\documentclass[pra,twocolumn]{revtex4-1}
\usepackage{hyperref,graphicx,subfig,color,filecontents,soul,physics,mathrsfs,relsize,amsmath,colortbl,tabularx,stackengine,dcolumn,bm,float,amssymb,tikz,accents}

\usepackage{tikz}
\usepackage{pgfplots}
\usetikzlibrary{arrows}
\usepackage[utf8]{inputenc}
\usetikzlibrary{intersections, pgfplots.fillbetween}
\usetikzlibrary{decorations.pathmorphing}
\newcommand\encircle[1]{%
  \tikz[baseline=(X.base)] 
    \node (X) [draw, shape=circle, inner sep=0] {\strut #1};}
\usetikzlibrary{arrows.meta}

\definecolor{LightCyan}{rgb}{0.88,1,1}
\definecolor{corn}{rgb}{0.98, 0.93, 0.36}
\definecolor{pastelyellow}{rgb}{0.99, 0.99, 0.59}
\definecolor{Green}{rgb}{0,0.6,0.2}

\newcommand{\be}{\begin{equation}}
\newcommand{\ee}{\end{equation}}
\newcommand{\bea}{\begin{eqnarray}}
\newcommand{\eea}{\end{eqnarray}}
\newcommand*{\myeqref}[2][Eq.~]{%
  \hyperref[{#2}]{#1(\ref*{#2})}%
}
\def\equationautorefname#1#2\null{%
  Eq.#1(#2\null)%
}

\begin{document}
\title{Band Gap Engineering and Controlling Transport Properties of Single Photons in Periodic and Disordered Jaynes-Cummings Arrays}
\author{Tiberius Berndsen$^\ast$, Nishan Amgain}
\thanks{These authors have contributed equally to this work.}
\author{Imran M. Mirza}
\email{mirzaim@miamioh.edu}
\affiliation{Macklin Quantum Information Sciences, \\Department of Physics, Miami University, Oxford, Ohio 45056, USA}

%%===================================================%%
%%                Article Abstract                   %%
%%===================================================%%
\begin{abstract}
We theoretically study the single photon transport properties in periodic and position-disordered Jaynes-Cummings (or JC) arrays of waveguide-coupled microtoroidal ring resonators, each interacting with a single two-level quantum emitter. Employing the real-space formalism of quantum optics, we focus on various parameter regimes of cavity quantum electrodynamics (cQED) to gain better control of single photon propagation in such a many-body quantum optical setting. As for some of the key findings, we observe that the periodic setting leads to the formation of the band structure in the photon transmission spectra, which is most evident in the strong coupling regime of cQCD. However, under the resonant conditions with no losses, the application of Bloch's theorem indicates that the width of forbidden gaps can be altered by tuning the emitter-cavity coupling to small values. Moreover, in the disordered case, we find that the single photon transmission curves show the disappearance of band formation. However, spectral features originating from cQED interactions observed for single atom-cavity problem remain robust against weak-disordered conditions. The results of this work may find application in the study of quantum many-body effects in the optical domain as well as in different areas of quantum computation and quantum networking. 
\end{abstract}

%%===================================================%%
%%              Sec.I: Introduction                  %%
%%===================================================%%
\maketitle
\section{\label{sec:I} Introduction}
Arrays of coupled atom-cavity systems offer state-of-the-art testbeds for studying strongly correlated many-body physics in the optical domain. With recent experimental advancements (thanks to the fabrication of micro-cavities and tremendous progress in controlling atomic and optical systems \cite{thompson2013coupling}), new avenues have been opened to utilize these arrays as quantum simulators of many-body physics \cite{hartmann2008quantum}. Unlike electron-based many-body systems (studied in condensed matter physics) where the separation between the neighboring sites is very small (e.g., in Josephson junction arrays \cite{cataliotti2001josephson} and optical lattices \cite{jaksch1998cold}), in coupled cavity arrays, the distance between neighboring cavities can be adjusted relatively freely \cite{lepert2011arrays}. This feature allows one to observe individual local properties of atom-cavity subsystems and their overall collective behavior \cite{ruiz2014spontaneous}.

In addition to employing such atom-cavity arrays in the development of exotic light-matter quantum states \cite{saxena2023realizing,schetakis2013frozen}, such architectures have shown crucial applications in the field of quantum information storage, transfer, and manipulation. Without atoms, such ``empty" waveguide-coupled ring resonators can store and delay classical light pulses and single photons \cite{heebner2002slow, mirza2013single}. In the presence of atoms, for instance, even at a few body levels, two coupled atom-cavity systems with a single excitation can exhibit many exciting phenomena of pure quantum nature, such as quantum state transfer and entanglement distribution \cite{cirac1997quantum, blais2020quantum, mirza2015bi}. The many-body extension to these systems generates entangled states of qubits \cite{bostelmann2023multipartite, mendoncca2020generation}, which can be robust to environmental losses by applying driven dissipative techniques \cite{mirza2022dissipative, stannigel2012driven}. Since the actual quantum networking protocols are believed to be made of many such atom-cavity structures, the study of the transport of single photons in such setups becomes a subject of immense value from the perspective of quantum information science \cite{meher2022review, baum2022effect}.

A literature review on the subject of single photon propagation in coupled cavity arrays with an on-site Jaynes-Cummings-like interaction reveals that in the past decades, such architectures have been studied for several quantum and photonic applications. For instance, Qin et al. have examined coupled cavity arrays in building quantum photonic switches \cite{qin2016controllable} and studied photon bound formation \cite{liao2010controlling}. Additionally, the existence of singularities in the photon transfer in the ultrastrong coupling regime of cQED \cite{felicetti2014photon}, controlling interactions between two atoms by altering the coupling rate between two cavities \cite{ogden2008dynamics}, optimization of quantum state transfer in Jaynes-Cummings-Hubbard models \cite{baum2022effect} and the emergence of photonic rogue waves in the dispersive regime of atom-cavity arrays \cite{cheng2022photonic} have also been reported. On the experimental side, the demonstration of strong coupling between an optical cavity with an atomic array \cite{liu2023realization} and the building of ultra-high quality factor microresonator arrays in wavelength-sized cavities have been reported \cite{notomi2008large} paving the way to achieve strong coupling regime of cQED.

Recently, we and others have studied single photon band gap properties in many-emitter waveguide quantum electrodynamics architectures and 
coupled-resonator optical waveguide chirally coupled with an array of two-level quantum emitters \cite{tang2022nonreciprocal, berndsen2023electromagnetically, sheremet2023waveguide}. In this work, by going beyond the periodic and chiral arrangements, we examined the single photon transport properties in periodic and disordered Jaynes-Cummings arrays with bi-directional or non-chiral on-site coupling between the atom and the cavity field. The main goal here is to study the interplay between the strong and weak cQED coupling regime and the periodic and disordered arrangements of the cavities on the single photon transport properties. Consequently, we study both atom-cavity arrays at a few-body level (up to ten waveguide-coupled atom-cavity subsystems either periodically arranged or with a position disorder in the location of cavities) and infinitely long extension of such an array.

As some of the key findings, in the strong coupling regime of cQED, we observe the formation of band gaps for ten atom-cavity arrays superimposed on the frequency doublet originating from the process of Rabi splitting. Taking the infinitely long extension of the array allowed us to apply Bloch's theorem, which, under no loss conditions, indicated the formation of forbidden frequency bands, which can be tuned by changing the atom-cavity detuning and coupling rate between the atom and cavity field modes. Finally, we briefly examine the situation in which the location of the cavities (ring resonators in our model) is disordered following a Gaussian distribution. The disorder destroys the band formation for arrays consisting of ten atom-cavity subsystems; however, the spectral features originating from the strong coupling regime of cQED remain more or less robust against weak disorder. 

The rest of the paper is organized as follows. In the next section, i.e., Sec.~\ref{sec:II}, we provide a theoretical description of this paper, including Hamiltonian, quantum state, and the equations obeyed by the probability amplitudes. In Sec.~\ref{sec:III}, we set the stage with a single emitter-cavity problem and discuss the single photon transport properties in various parameter regimes of cQED. Next, in Sec.~\ref{sec:IV}, we examine the photon transmission in a periodic lattice of many (up to ten) emitter-cavity subsystems. We also take the infinitely many emitter-cavity limits in the same section to discuss the dispersion properties of such lattices. Following that, in Sec.~\ref{sec:V}, we briefly examine the disordered emitter-cavity lattices to analyze the impact of the position disorder in the location of ring resonators on the photonic transport properties. Finally, in Sec.~\ref{sec:VI}, we close with this work's main conclusions and indicate some possible future directions. 

%%===================================================%%
%%                JC Array Setup                     %%
%%===================================================%%
%\begin{figure*}[t]
%\includegraphics[width=6.8in,height=1.95in]%{Fig1}
%\captionsetup{
% format=plain,
% margin=1em,
% justification=raggedright,
% singlelinecheck=false
%}
% \caption{(Color online) A 1D chain of fiber coupled atom-cavity system: JC array. The chain is divided into $N$ segments where each segment consistes of an atom-cavity system coupled with a fiber of lebgth $L$. A single photon launched from $x=-\infty$ propagates through this array where reflection and transmission intensities from any $i$th segment are respectively represented by $R_{i}$ and $T_{i}$.}\label{Fig1}
%\end{figure*}
%%%%%%%%%%%%%%%%%%%%%%%%%%%%%%%%%%%%%%%%%%%%%%%%%%%%%%%%%%%
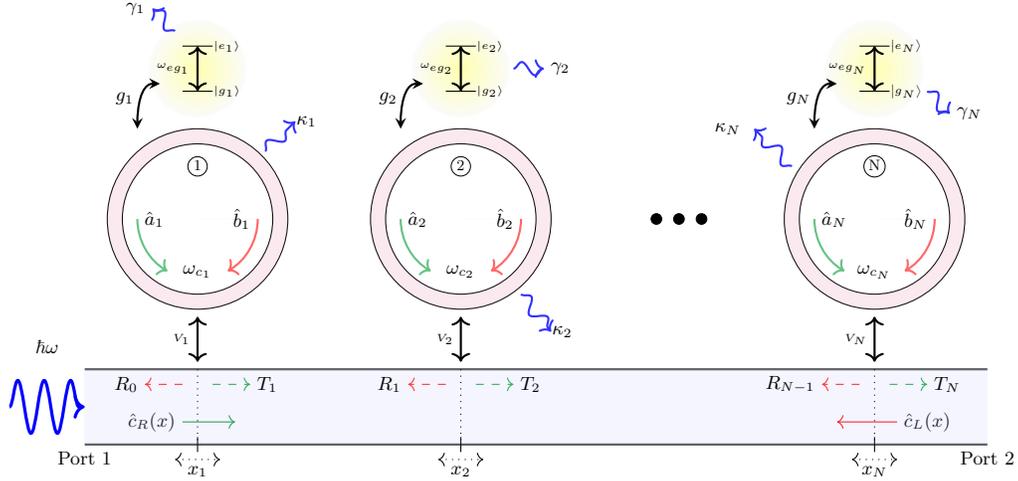
\begin{figure*}
\centering
%%%%%%%%%%%%%%%%%%%%%%%tikzpicture 1%%%%%%%%%%%%%%%%%%%%%%%%%%
    \begin{tikzpicture}[every node/.style = {scale=0.8}]
    %drawing the waveguides
    %lower waveguides
    \draw [black!70, thick, name path = C](0,-5)  -- (12,-5);
    \draw [black!70, thick, name path = D](0,-6)node [color = black, anchor = north] {Port 1} -- (12,-6)node[color = black, anchor = north] {Port 2};
    \draw [->,Green] (1.3,-5.7) node [anchor = east, black] {$\hat{c}_R(x)$} --(2.0,-5.7);
    \draw [->,red] (10.8,-5.7) node [anchor = west, black] {$\hat{c}_L(x)$} --(10.0,-5.7);
    %filling the area between the waveguides
    \tikzfillbetween[of= C and D]{blue!20, opacity=0.1};
    \draw (1.5, -5.9) -- (1.5,-6.1);
    \draw (5.0, -5.9) -- (5.0,-6.1);
    \draw (10.5,-5.9) -- (10.5,-6.1);
    %filling the area between the waveguides
    \tikzfillbetween[of= C and D]{blue!20, opacity=0.1};
%photon
    \draw[->, blue, very thick, decorate, decoration={snake, amplitude = 10}] (-1,-5.5) -- (0,-5.5);
    \node at (-0.5, -4.7) {$\hbar\omega$};
    
%ring resonators
    
    \draw [name path = c1inside] (1.5,-3) circle (1.2);
    \draw [name path = c1outside](1.5,-3) circle (1.0);
    \tikzfillbetween[of= c1inside and c1outside]{purple!40, opacity=0.2};
    \draw[->, thick, Green!60] (0.7,-3)node[anchor = west, color = black] {$\hat{a}_1$} arc (180:240:0.8);
    \draw[->, thick, red!60](2.3,-3)node[anchor = east, color = black] {$\hat{b}_1$} arc (-180:-240:-0.8);
    \node at (1.5,-2.3) {\encircle{1}};
    \node at (1.5,-3.7) {$\omega_{c_1}$};
    
    \draw [name path = c2inside] (5,-3) circle (1.2);
    \draw [name path = c2outside] (5,-3) circle (1.0);
    \tikzfillbetween[of= c2inside and c2outside]{purple!40, opacity=0.2};
    \draw[->, thick, Green!60] (4.2,-3)node[anchor = west, color = black] {$\hat{a}_2$} arc (180:240:0.8);
    \draw[->, thick, red!60](5.8,-3)node[anchor = east, color = black] {$\hat{b}_2$} arc (-180:-240:-0.8);
    \node at (5.0,-2.3) {\encircle{2}};
    \node at (5.0,-3.7) {$\omega_{c_2}$};

    \filldraw [black] (7.6,-3) circle (2pt)   (7.9,-3) circle (2pt)   (8.2,-3) circle (2pt);
    
    \draw [name path = c3inside] (10.5,-3) circle (1.2);
     \draw [name path = c3outside] (10.5,-3) circle (1.0);
     \tikzfillbetween[of= c3inside and c3outside]{purple!40, opacity=0.2};
     \draw[->, thick, Green!60] (9.7,-3)node[anchor = west, color = black] {$\hat{a}_N$} arc (180:240:0.8);
    \draw[->, thick, red!60](11.3,-3)node[anchor = east, color = black] {$\hat{b}_N$} arc (-180:-240:-0.8);
    \node at (10.5,-2.3) {\encircle{N}};
    \node at (10.5,-3.7) {$\omega_{c_N}$};
%atoms
    \shade[shading=radial, outer color=gray!5, inner color = yellow!60, opacity = 0.5] (1.5,-1.0) circle (0.65);
    \shade[shading=radial, outer color=gray!5, inner color = yellow!60, opacity = 0.5] (5,-1.0) circle (0.65);
    \shade[shading=radial, outer color=gray!5, inner color = yellow!60, opacity = 0.5] (10.5,-1.0) circle (0.65);
%drawing the ground and excited state of the emitters
    \draw   (1.3,-0.7) -- (1.7,-0.7) node[anchor = west, shift={(-0.1,0)}]            {\tiny $\ket{e_1}$}
            (1.3,-1.3) -- (1.7,-1.3) node[anchor = west, shift={(-0.1,0)}] {\tiny $\ket{g_1}$}
            (4.8,-0.7) -- (5.2,-0.7) node[anchor = west, shift={(-0.1,0)}] {\tiny $\ket{e_2}$}
            (4.8,-1.3) -- (5.2,-1.3) node[anchor = west, shift={(-0.1,0)}] {\tiny $\ket{g_2}$}
            (10.3,-0.7) -- (10.7,-0.7) node[anchor = west, shift={(-0.1,0)}] {\tiny $\ket{e_N}$}
            (10.3,-1.3) -- (10.7,-1.3) node[anchor = west, shift={(-0.1,0)}] {\tiny $\ket{g_N}$};
%atomic resonance frequency
    \draw[<->, thick] (1.5, -0.7) -- (1.5,-1.0)node[anchor = east] {\tiny $\omega_{eg_1}$} -- (1.5,-1.3);
    \draw[<->, thick] (5.0, -0.7) -- (5.0,-1.0)node[anchor = east] {\tiny $\omega_{eg_2}$} -- (5,-1.3);
    \draw[<->, thick] (10.5, -0.7) -- (10.5,-1.0)node[anchor = east] {\tiny $\omega_{eg_N}$} -- (10.5,-1.3);
%drawing the coupling arrows ring to waveguide
    \draw[<->, thick] (1.5, -4.3) -- (1.5,-4.6)node[anchor = east, black] {\tiny $V_1$} -- (1.5,-4.9);
    \draw[<->, thick] (5.0, -4.3) -- (5.0,-4.6)node[anchor = east, black] {\tiny $V_2$} -- (5,-4.9);
    \draw[<->, thick] (10.5, -4.3) -- (10.5,-4.6)node[anchor = east, black] {\tiny $V_N$} -- (10.5,-4.9);
%drawing the coupling arrows atom to ring
    \draw[<->,>=stealth, thick,out = 90, in = 180](0.7,-1.8)  to node [anchor = east, midway] {$g_1$} (1.0,-1.2);
    \draw[<->,>=stealth, thick,out = 90, in = 180](4.2,-1.8)  to node [anchor = east, midway] {$g_2$} (4.5,-1.2);
    \draw[<->,>=stealth, thick,out = 90, in = 180](9.7,-1.8)  to node [anchor = east, midway] {$g_N$} (10.0,-1.2);
%spontanenous emission for ring resonators
    \draw[->,blue!80, thick, decorate, decoration={snake}] (2.4,-2.1) -- (2.8,-1.7) node[black, anchor=west, shift={(-0.1,0)}]{$\kappa_{1}$};
    \draw[->,blue!80, thick, decorate, decoration={snake}] (5.8,-4) -- (6.2,-4.5)node[black, anchor=west, shift={(-0.1,0)}]{$\kappa_{2}$};
    \draw[->,blue!80, thick, decorate, decoration={snake}] (9.4,-2.3) -- (8.9,-1.8) node[black, anchor=east, shift={(-0.1,0)}]{$\kappa_{N}$};
%spontaneous emission for atoms
  \draw[->,blue!80, thick, decorate, decoration={snake}] (1.2,-0.5) -- (0.9,-0.2) node[black, anchor = east]{$\gamma_{1}$};
  \draw[->,blue!80, thick, decorate, decoration={snake}] (5.7,-1.0) -- (6.1,-1.0) node[black, anchor = west]{$\gamma_{2}$};
  \draw[->,blue!80, thick, decorate, decoration={snake}] (11.2,-1.3) -- (11.5,-1.6) node[black, anchor = west]{$\gamma_{N}$};
%random oscillations in position of ring resonator
\draw[dotted, <->] (1.2,-6.2) --(1.5,-6.2) node [anchor = north] {$x_1$} -- (1.8,-6.2);
\draw[dotted, <->] (4.7,-6.2) --(5.0,-6.2) node [anchor = north] {$x_2$} -- (5.3,-6.2);
\draw[dotted, <->] (10.2,-6.2) --(10.5,-6.2) node [anchor = north] {$x_N$} -- (10.8,-6.2);

%Reflections and Transmissions
\draw[dotted] (1.5,-5)--(1.5,-6);
\draw[dotted] (5.0,-5)--(5.0,-6);
\draw[dotted] (10.5,-5)--(10.5,-6);
    %Reflections
    \draw[->, red!80, dashed] (1.3,-5.2) -- (0.8, -5.2) node[anchor = east, black] {$R_0$};
    \draw[->, red!80, dashed] (4.8,-5.2) -- (4.3, -5.2) node[anchor = east, black] {$R_1$};
    \draw[->, red!80, dashed] (10.3,-5.2) -- (9.8, -5.2) node[anchor = east, black] {$R_{N-1}$};
     %Transmissions
    \draw[->, Green!80, dashed] (1.7,-5.2) -- (2.2, -5.2) node[anchor = west, black] {$T_1$};
    \draw[->, Green!80, dashed] (5.2,-5.2) -- (5.7, -5.2) node[anchor = west, black] {$T_2$};
    \draw[->, Green!80, dashed] (10.7,-5.2) -- (11.2, -5.2) node[anchor = west, black] {$T_{N}$};

%wiggles
    
    %gaussian wiggles?
    %\shade[shading=radial, outer color=black!5, inner color = black!80]  (1,0.2) ellipse (0.4 and 0.05); \node at (1,0.4) {\scriptsize $x_1$};
    %\shade[shading=radial, outer color=black!5, inner color = black!80]  (3,0.2) ellipse (0.4 and 0.05); \node at (3,0.4) {\scriptsize $x_2$};
    %\shade[shading=radial, outer color=black!5, inner color = black!80]  (6,0.2) ellipse (0.4 and 0.05); \node at (6,0.4) {\scriptsize $x_N$};
    
\end{tikzpicture}
\captionsetup{
  format=plain,
  margin=1em,
 justification=raggedright,
 singlelinecheck=false
}
\caption{(Color online) A 1D chain of waveguide coupled atom-cavity system known as the Jaynes-Cummings array. The chain is divided into $N$ segments, each consisting of an atom-cavity subsystem coupled with a common bidirectional waveguide. A single photon is launched from the left end of the waveguide (from Port 1). The reflection and transmission intensities from any $i$th segment are represented by $R_{i}$ and $T_{i}$. The lattice constant (separation between two consecutive cavities) is defined by $x_{j+1}-x_j=L$. At Port 1 and 2, the net reflection probability $R_0$ and net transmission probability $T_N$ are recorded, respectively.}\label{Fig1}
\end{figure*}
%%%%%%%%%%%%%%%%%%%%%%%%%%%%%%%%%%%%%%%%%%%%%%%%%%%%%%%%

%%===================================================%%
%%          Sec.II: Theortical Description           %%
%%===================================================%%
\section{\label{sec:II} Theoretical Description}
\subsection{\label{sec:IIA} Model Hamiltonian}
The system under consideration consists of a one-dimensional atom-cavity array coupled through a lossless bidirectional waveguide (or optical fiber), as depicted in Fig.~\ref{Fig1}. We employ the real-space quantization technique as initially introduced by Anderson in the 1960s \cite{anderson1961localized, wiegmann1983exact} and more recently applied by Fan and others in problems related to waveguide quantum electrodynamics \cite{shen2005coherent,shi2011two}, cQED \cite{shen2009theoryI,shen2009theoryII}, circuit QED \cite{chen2014scattering}, cavity optomechanics \cite{ren2013single}, hybrid atom-optomechanics \cite{jia2013single} and quantum plasmonics \cite{chen2011surface}. The total Hamiltonian of the system $\hat{\mathcal{H}}$ can be decomposed into six parts:
\begin{equation}\label{eq:Hsys}
\hat{\mathcal{H}}=\hat{\mathcal{H}}_{a}+\hat{\mathcal{H}}_{c}+\hat{\mathcal{H}}_{w}+\hat{\mathcal{H}}_{ac}+\hat{\mathcal{H}}_{cc}+\hat{\mathcal{H}}_{cw}.
\end{equation} 
Here $\hat{\mathcal{H}}_{a}$, $\hat{\mathcal{H}}_{c}$, $\hat{\mathcal{H}}_{w}$, $\hat{\mathcal{H}}_{ac}$, $\hat{\mathcal{H}}_{cc}$, and $\hat{\mathcal{H}}_{cw}$ represent the atom, cavity, waveguide, atom-cavity interaction, cavity backscattering, and cavity-waveguide interaction part of the Hamiltonian, respectively. In $\hat{\mathcal{H}}_{a}$, we are considering atoms as qubits with $j$th atom excited (ground) state $\ket{e_{j}}(\ket{g_{j}})$ with the lowering operator, raising operator, transition frequency, and spontaneous emission rate given by $\hat{\sigma}_j$, $\hat{\sigma}^\dagger_j$, $\omega_{eg_j}$ and $\gamma_{j}$, respectively $(\forall j = 1,2,3,..., N)$. The explicit form of $\hat{\mathcal{H}}_a$ is given by
\begin{align}
\hat{\mathcal{H}}_a=\hbar\sum^N_{j=1}\widetilde{\omega}_{eg_{j}}\hat{\sigma}^\dagger_j\hat{\sigma}_j,
\end{align}
where $\widetilde{\omega}_{eg_{j}}\equiv\omega_{{eg}_j}-i\gamma_j$. The optical cavities in our model are microtoroidal ring resonators \cite{spillane2005ultrahigh,armani2003ultra} where for any $j$th cavity in the array, destruction of a photon in clockwise (counter-clockwise) cavity mode is described by annihilation operator $\hat{b}_j (\hat{a}_j)$. In contrast, both modes have the same resonant frequency $\omega_{c_j}$. The free Hamiltonian for the cavities $\hat{\mathcal{H}}_{c}$ takes the form
\begin{align}
    \hat{\mathcal{H}}_c = \hbar\sum^N_{j=1}\widetilde{\omega}_{c_j}\left(\hat{a}^\dagger_j\hat{a}_j+\hat{b}^\dagger_j\hat{b}_j\right).
\end{align}
where we have neglected the zero-point energies, with $\kappa_{j}$, incorporated in the modified cavity frequency through $\widetilde{\omega}_{c_j} \equiv\omega_{c_j}-i\kappa_j$, is the photon leakage rate. Depending on the propagation direction of the photon in the waveguide, we define two position-dependent annihilation operators: $\hat{c}_{L}(x)$ and $\hat{c}_{R}(x)$ where $L/R$ stands for left/right propagation direction in the waveguide. In the real-space formalism of quantum optics the waveguide Hamiltonian $\hat{\mathcal{H}}_{w}$ thus takes the following form
\begin{align}
    \hat{\mathcal{H}}_{w}&=\hbar\int_{-\infty}^{\infty}\hat{c}^{\dagger}_{R}(x)\left(\omega_{0}-iv_{g}\frac{\partial}{\partial x}\right)\hat{c}_{R}(x)dx\nonumber\\
    &+\hbar\int_{-\infty}^{\infty}\hat{c}^{\dagger}_{L}(x)\left(\omega_{0}+iv_{g}\frac{\partial}{\partial x}\right)\hat{c}_{L}(x)dx,
\end{align}
with $\omega_0$ being the frequency around which the waveguide dispersion relation has been linearized (which can be set to zero without loss of generality). $v_{g}$ is the group velocity of the photon in the waveguide. In the interaction parts of the Hamiltonian, the first term $\hat{\mathcal{H}}_{ac}$ shows the Jaynes Cummings interaction between each mode in a given ring cavity with its respective atom. The strength of interaction between cavity mode $\hat{a}_{i} (\hat{b}_{i})$ with respective atom is represented by $g_{a_j} (g_{b_j})$ which is set equal to $g_j$ for simplicity. With these considerations one can express the $\hat{\mathcal{H}}_{ac}$ as
\begin{align}
    \hat{\mathcal{H}}_{ac}=\hbar \sum^N_{j=1} \left(g_{j}\hat{a}_j\hat{\sigma}^{\dagger}_j + g_{j} \hat{b}^{\dagger}_j\hat{\sigma}_j +h.c. \right),
\end{align}
where $h.c.$ represents terms that are Hermitian conjugate to the first two terms in the parenthesis; note that the atomic location near the circumference of the ring cavity has been ignored in our model. Next, the cavity backscattering Hamiltonian $\hat{\mathcal{H}}_{cc}$ models the interaction between the two cavity modes in the $j$th cavity with the strength $\eta_{j}$. This part of the interaction Hamiltonian can be modeled as
\begin{align}
\hat{\mathcal{H}}_{cc}=\hbar\sum^N_{i=1}\left(\eta_j\hat{a}_j\hat{b}^\dagger_j + h.c. \right).
\end{align}
Finally, we point out that in our model, upon entering the ring cavity, the photon can escape into the waveguide due to the mechanism of evanescent cavity-waveguide coupling represented by the parameter $V_j$ for the $jth$ cavity. Thus, under the rotating wave approximation, the cavity-waveguide interaction Hamiltonian can be expressed as
\begin{align}
    \hat{\mathcal{H}}_{cw} & = \hbar\sum^N_{j=1}\Big(V_j\hat{c}^{\dagger}_{R}(x_j)\hat{a}_j + V^\ast_j\hat{c}^{\dagger}_{L}(x_j)\hat{b}_j + h.c. \Big).
\end{align}
For simplicity, we set $V_{a_j}=V_{b_j}=V_{j}$, which describes the interaction between the waveguide and the j$^{th}$ cavity modes. The dependence of $\hat{c}^\dagger_{R/L}$ (and consequently of $\hat{c}_{R/L}$) on $x_j$ represents the location of this interaction occurring at position $x_j$ for the j$^{th}$ cavity. The non-vanishing commutation relations among various system operators are summarized as
\begin{equation*}
\begin{split}
&\lbrace\hat{\sigma}^{\dagger}_{j},\hat{\sigma}_{l}\rbrace=\delta_{jl},\hspace{2mm}[\hat{a}_{j},\hat{a}^{\dagger}_{j^{'}}]=\delta_{jj^{'}},\hspace{2mm}[\hat{b}_{j},\hat{b}^{\dagger}_{j^{'}}]=\delta_{jj^{'}},\\
&[\hat{c}_{\alpha}(x),\hat{c}^{\dagger}_{\beta}(x^{'})]=\delta_{\alpha\beta}\delta(x-x^{'}), ~\forall \alpha=L, R; ~\forall \beta=L, R.
\end{split}
\end{equation*}
Finally, as shown in Fig.~\ref{Fig1}, we note that due to the geometry of our setup, the $\hat{a}_j$ mode  ($\hat{b}_j$ mode) is coupled to the right (left) propagation direction in the waveguide. 
%%%%%%%%%%%%%%%%%%%%%%%%%%%%%%%%%%%%%%%%%%%%%%%%%%%%%%%%

\subsection{\label{sec:IIB} Quantum state and the amplitude equations}
To obtain the single photon transport properties in the system under study, we perform a stationary analysis of the problem. To this end, we take the following form of the state of the system restricted to the single-excitation sector of the Hilbert space
\begin{align}\label{eq:Psi}
    \ket{\Psi} = &\Bigg[\sum_{\alpha=L,R}\int_{-\infty}^{\infty} \varphi_{\alpha}(x)\hat{c}^{\dagger}_{\alpha}(x)dx\nonumber\\
    &+\sum^N_{j=1}\Big(e_{q_j}\hat{\sigma}^\dagger_j+ e_{a_j}\hat{a}^\dagger_j + e_{b_j}\hat{b}^\dagger_j\Big) \Bigg]\ket{\varnothing},
\end{align}
where $\varphi_{L/R}(x)$ is the single-photon wavefunction of the left/right moving continuum in the waveguide and $e_{q_j}, e_{a_j}$, $e_{b_j}$ are the probability amplitudes of finding the $j$the atom/qubit excited, one photon in the counterclockwise mode of the $j$th ring cavity, and one photon in the clockwise mode of the $j$th ring cavity, respectively. $\ket{\varnothing}$ represents the system's ground state with no photons in any of the ring cavities, all atoms in their ground state, and no photons in the waveguide. 

Inserting Eq.~(\ref{eq:Psi}) along with the total Hamiltonian specified in Eq.~(\ref{eq:Hsys}) in the time-independent Schr\"odinger equation $\hat{\mathcal{H}}\ket{\Psi}=\hbar\omega\ket{\Psi}$ we obtain the following set of equations obeyed by the probability amplitudes
\begin{subequations}
\label{eq:TIAE}
\begin{eqnarray}
-iv_{g}\frac{\partial \varphi_{R}(x)}{\partial x}+\sum^N_{j=1}e_{a_j}V_{j}\delta(x-x_j)=(\omega-\omega_{0})\varphi_{R}(x),~~\\
+iv_{g}\frac{\partial \varphi_{L}(x)}{\partial x}+\sum^N_{j=1}e_{b_j}V_{j}\delta(x-x_j)=(\omega-\omega_{0})\varphi_{L}(x),~~\\
V^{\ast}_{j}\varphi_{R}(x_j)+g^{\ast}_{j}e_{q_j}+\eta_j e_{b_j}=(\omega-\widetilde{\omega}_{c_j})e_{a_j},~~\\
V^{\ast}_{j}\varphi_{L}(x_j)+g^{\ast}_{j}e_{q_j}+\eta^\ast_j e_{a_j}=(\omega-\widetilde{\omega}_{c_j})e_{b_j},~~\\
g_{j}e_{a_j}+g_{j}e_{b_j}=(\omega-\widetilde{\omega}_{{eg}_j})e_{q_j}.~~
\end{eqnarray}
\end{subequations}
Note that $\hbar\omega$ is the energy of the single photon launched from the left end of the waveguide, and the last three subequations are, in fact, a system of linear equations depending on the number of atom-cavity subsystems $N$ as $1\leq j \leq N$. In the next section, we show how the above system of equations can be solved for a single (i.e., $N=1$) atom-cavity case, which will form the basis of our calculations for the larger $N$ value problems.

\section{\label{sec:III} Single atom-cavity setup}
\begin{figure*}[t]
\centering
  \begin{tabular}{@{}cccc@{}}
    \includegraphics[width=1.65in, height=1.35in]{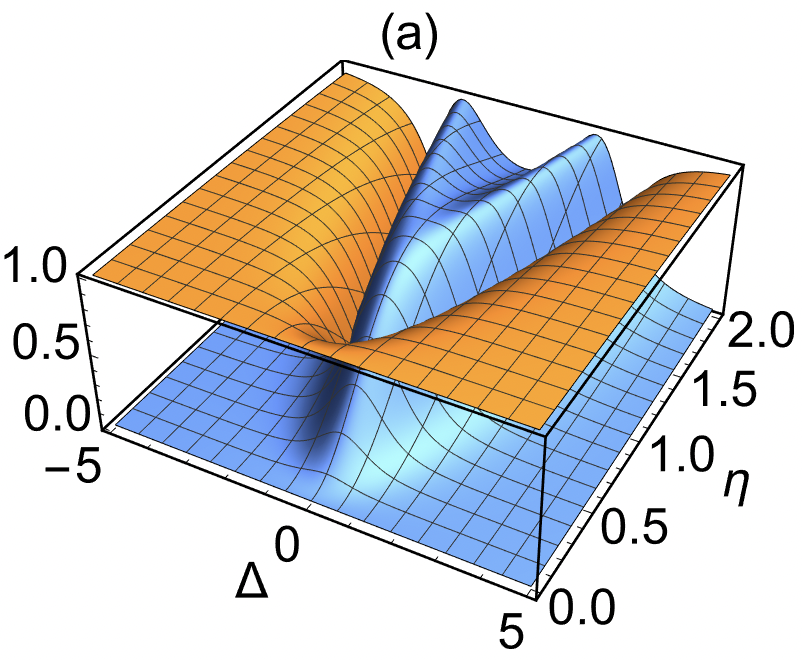} &
    \includegraphics[width=1.65in, height=1.35in]{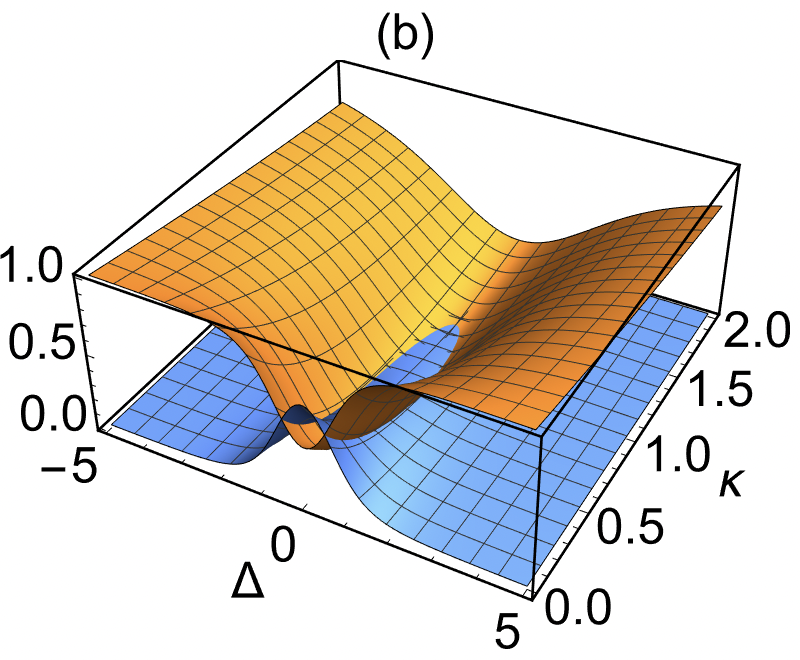} &
    \includegraphics[width=1.65in, height=1.35in]{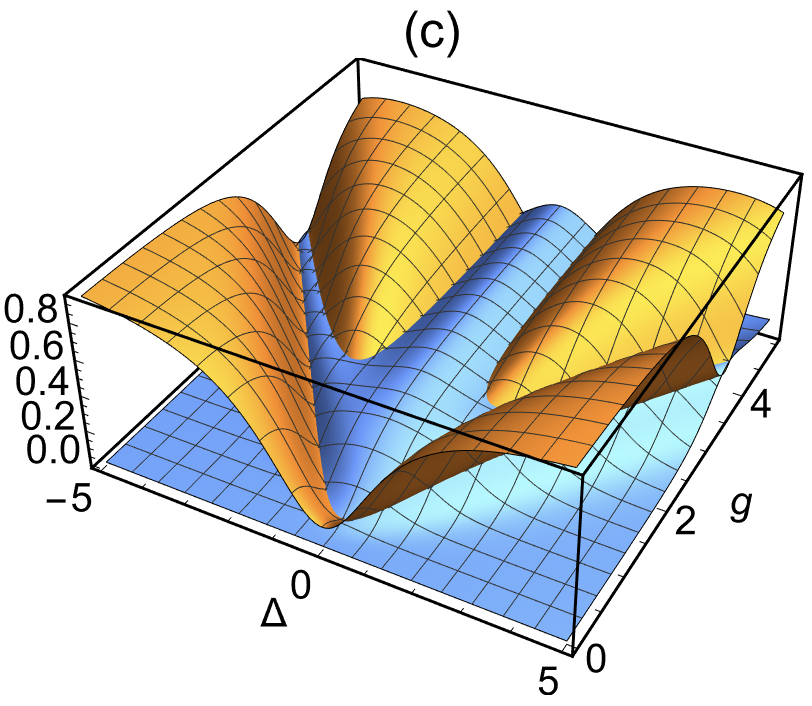} \\
    \includegraphics[width=1.65in, height=1.35in]{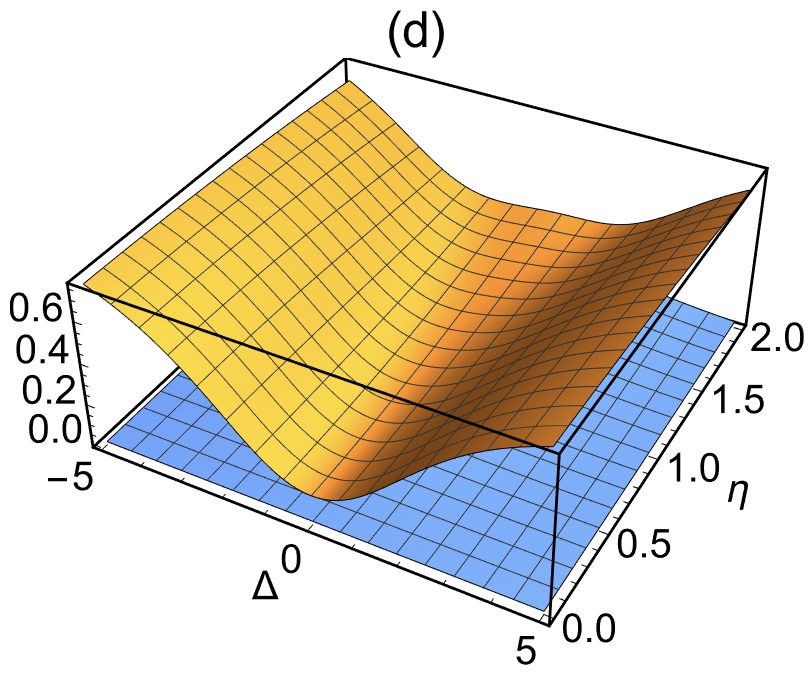} &
    \includegraphics[width=1.65in, height=1.35in]{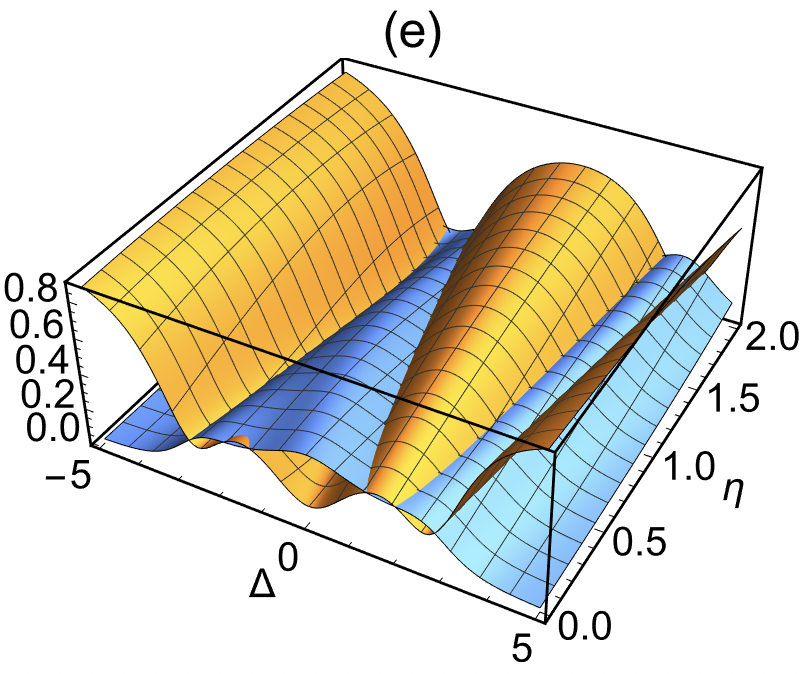} &
    \includegraphics[width=1.65in, height=1.35in]{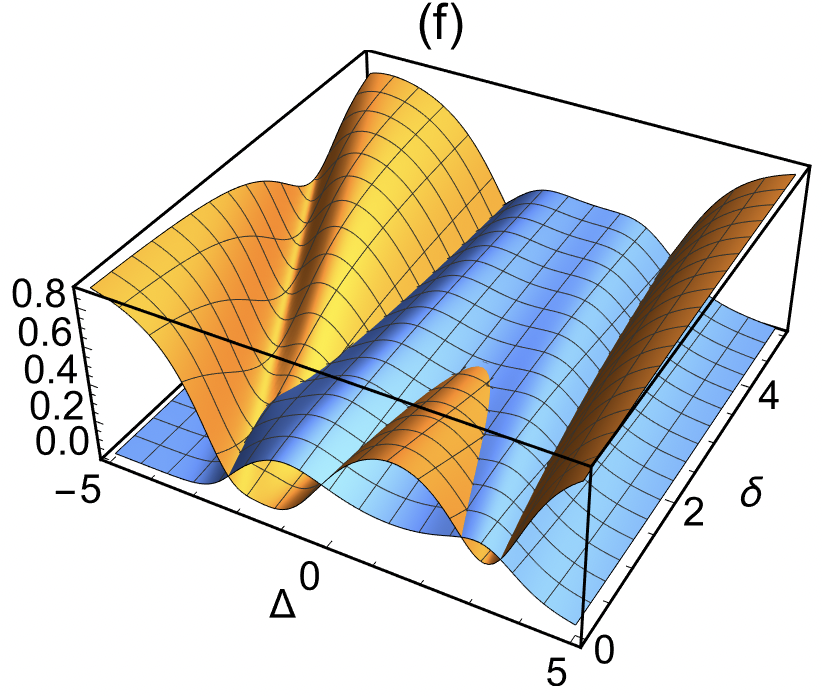} 
  \end{tabular}
  \captionsetup{
  format=plain,
  margin=1em,
 justification=raggedright,
 singlelinecheck=false
}
\caption{(Color online) Single-photon transport through a single atom-cavity-waveguide system is described in terms of reflected (blue curves) and transmission spectra (golden curves). In all plots, $\Gamma$ has been used as the unit of the problem, while for plots (a) to (e), we consider no detuning between atomic transition frequency and cavity resonance. {\bf(a)} Case-I: Atom decoupled with no loss. We set $g=0, \kappa=\gamma=0$ and varied cavity backscattering $\eta= 0 \rightarrow 2\Gamma$. {\bf(b)} Case-II: Atom decoupled with losses. All parameters are the same as in case-I while cavity leakage rate $\kappa$ has been varied from $0$ to $2\Gamma$. {\bf((c)} Case-III: Weak to strong coupling regime with no backscattering. In this case we fixed $\eta=0$ and $\kappa=\gamma=0.5\Gamma$ while changing $g$ from $0$ to $5\Gamma$. {\bf((d)} Case-IV: Weak coupling with backscattering. For this case, we selected $g=0.5\Gamma, \kappa=\gamma=2\Gamma$. Backscattering rate $\eta$, in this and next case, has been changed between $0$ and $2\Gamma$. {\bf((e)} Case-V: Strong coupling with backscattering. $g=2\Gamma$, $\kappa=\gamma=0.5\Gamma$. {\bf((f)} Case-VI: Strong coupling with atom-cavity detuning. $g=2\Gamma$, $\kappa=\gamma=0.5\Gamma$, $\eta=1\Gamma$ and $\Delta_{ac}\equiv \omega_{c}-\omega_{eg}=\delta$ varied from no detuning to large detuning of $5\Gamma$.}\label{Fig2}
\end{figure*}

To obtain the reflection probability $R$ and transmission probability $T$ (which in turn are related to their respective reflection and transmission coefficients through $R=|r|^2$ and $T=|t|^2$), for the single atom-cavity problem, we take the following ansatz for the waveguide wavefunctions
\begin{equation}
\label{eq:phis}
\begin{split}
&\phi_{R}(x)=e^{ikx}\left[\Theta(-x)+t\Theta(x)\right],\\
&\phi_{L}(x)=re^{-ikx}\Theta(-x)
\end{split}
\end{equation}  
Here $\Theta(\pm x)$ is the Heaviside step function. The parameter $k$ (which has the units of inverse length) is related to the relevant frequencies of the problem through $k v_{g}=\omega-\omega_{0}$ with $\omega\equiv\omega_{0}\pm k_{R/L}v_{g}$ and $k_{R/L}$ being the rightward/leftward waveguide mode wavenumber. Thus, inserting Eq.~(\ref{eq:phis}) into Eq.~(\ref{eq:TIAE}) we arrive at the following set of single-photon transport equations
\begin{subequations}
\label{eq:TIAEtr}
\begin{eqnarray}
-iv_{g}(t-1)+e_{a}V = 0\\
-iv_{g}r+e_{b}V = 0\\
-\widetilde{\Delta}_{c}e_{a}+V^{\ast}\left(\frac{1+t}{2}\right)+g e_{q}+\eta e_{b}=0\\
-\widetilde{\Delta}_{c}e_{b}+V^{\ast}\left(\frac{r}{2}\right)+g^{\ast} e_{q}+\eta^{\ast} e_{a}=0\\
-\widetilde{\Delta}_{eg}e_{q}+g^{\ast}e_{a}+ge_{b}=0,
\end{eqnarray}
\end{subequations}
where we have adopted a short notation in which  $\widetilde{\Delta}_{c}\equiv\omega-\widetilde{\omega}_{c}$ and $\widetilde{\Delta}_{eg}\equiv\omega-\widetilde{\omega}_{eg}$. Solving the above set of equations for $r$ and $t$ variables yields the desired transmission and reflection amplitudes as
\begin{equation}
\begin{split}
&t=\frac{\widetilde{\Delta}_{c}\left(\widetilde{\Delta}_{eg}\widetilde{\Delta}_{c}-2|g|^{2}\right)+\widetilde{\Delta}_{eg}\left(\Gamma^{2}-|\eta|^2\right)-2|g|^{2}\eta}{\left(\widetilde{\Delta}_{c}+i\Gamma\right)\left\lbrace\widetilde{\Delta}_{eg}(\widetilde{\Delta}_{c}+i\Gamma)-2|g|^{2}\right\rbrace-2|g|^{2}\eta-|\eta|^{2}\widetilde{\Delta}_{eg}},\\
&r=\frac{-2i\Gamma\left(\widetilde{\Delta}_{eg}\eta+|g|^2 \right)}{\left(\widetilde{\Delta}_{c}+i\Gamma\right)\left\lbrace\widetilde{\Delta}_{eg}(\widetilde{\Delta}_{c}+i\Gamma)-2|g|^{2}\right\rbrace-2|g|^{2}\eta-|\eta|^{2}\widetilde{\Delta}_{eg}}.
\end{split}
\end{equation}
In the above set of equations, we introduce the parameter $2\Gamma=V^{2}/v_{g}$, which characterizes the cavity-waveguide coupling strength and contributes to defining the line width of the intracavity modes. In Fig.~\ref{Fig2}, we plot the single-photon transmission $T$ and the reflection $R$ spectra. There are various parameters involved in the system dynamics, which can be selected in many different ways to plot $T$ and $R$ (for a detailed account at the level of single atom-cavity-waveguide setup, see \cite{srinivasan2007mode,shen2009theoryII}). Here, we focus on six cases of interest, as discussed separately below.

\hspace{-3.5mm}$\bullet$ {\bf Decoupled atom case with no losses}: We start with the simplest possible case in Fig.~\ref{Fig2}(a) in which we decouple the atom from our atom-cavity-waveguide subsystem completely, i.e., we set $g=0$ and also ignore all losses (no spontaneous emission and no photon leakage from the cavity). Here (and for the first five cases discussed below), we consider an on-resonance case in which $\omega_{eg}=\omega_c$ or $\Delta_{ac}\equiv\omega_c-\omega_{eg}=0$ such that $\Delta_c=\Delta_{eg}\equiv\Delta$ and discuss the impact of cavity backscattering on transmission and reflection spectra. In this case, the transmission and reflection coefficients take a much-simplified form as
\begin{align}
t=-\frac{\Gamma^2+\Delta^2-\eta^2}{\left(\Gamma-i\Delta\right)^2+\eta^2}~~\textit{and}~~r=\frac{2i\Gamma\eta}{\left(\Gamma-i\Delta\right)^2+\eta^2}.
\end{align}
We begin by observing that if we set $\eta=0$, then our setup works as an all-pass filter, i.e., $\forall \Delta$, we obtain a 100\% transmission (golden curve) with null reflection (blue curve in Fig.~2) however, when $\eta$ increases, at $\Delta=0$, a finite reflection appears with a Lorentzian-type profile. Finally, when large backscattering (i.e., as $\eta$ approaches 1.5$\Gamma$ and higher), transmission and reflection spectra split up into two peaks around the resonant point. This behavior seems to mimic the emergence of frequency doublet spectra as discussed below in the strong coupling regime of cavity quantum electrodynamics \cite{kimble1998strong}. 

\hspace{-3.5mm}$\bullet$ {\bf Decoupled atom case with losses:} Next, in Fig.~\ref{Fig2}(b), we retain the atomic decoupling case but include the losses from the cavity and from atom into consideration. For simplicity, in this case we set $\gamma=\kappa$ and find that transmission and reflection amplitudes take the form
\begin{align}
    &t=1-\frac{\Gamma}{\Gamma-i(\Delta-\eta)+\kappa}-\frac{\Gamma}{\Gamma-i(\Delta+\eta)+\kappa}~\text{and}\nonumber\\
    & r=\frac{2i\Gamma\eta}{(\Gamma-i(\Delta-\eta)+\kappa)(\Gamma-i(\Delta+\eta)+\kappa)}
\end{align}
As the cavity loss rate $ \kappa $ increases (say from $ \kappa=\Gamma $ to $ \kappa=2\Gamma $), reflection tends to reach an almost null value around resonant. On the other hand, the transmission spectra exhibit an enhancement at the resonant point along with an increase in the transmission line width. 

\hspace{-3.5mm}$\bullet$ {\bf Weak to strong coupling regime with no backscattering}: In Fig.2(c), we begin from a weak coupling regime of cQED ($g<(\kappa,\gamma)$) in which both cavity loss rate and spontaneous emission rate are set to be equal to $0.5\Gamma$ and increase $g$ gradually to enter the strong coupling regime ($g>(\kappa,\gamma)$). Cavity backscattering has been turned off for this plot. For this case, we obtain 
\begin{align}
    &t=1-\frac{\Gamma}{\Gamma-i\Delta+\kappa}-\frac{\Gamma(\gamma-i\Delta)}{2g^2+(\gamma-i\Delta)(\Gamma-i\Delta+\kappa)},\nonumber\\
    & r=\frac{2g^2\Gamma}{(\Gamma-i\Delta+\kappa)(2g^2+(\gamma-i\Delta)(\Gamma-i\Delta+\kappa))}.
\end{align}
As before, when $g=0$, we find our setup to be a pass filter. However, as $g$ increases, within the weak coupling regime of cQED, we find that a single Lorentzian peak begins to form. Next, when $g$ is further increased, we enter the strong coupling regime. Here, we observe the formation of three dips (maxima) in the transmission (reflection) spectrum, as there are two counter-propagating modes in the ring cavity, which, in the absence of backscattering, form two standing waves in the cavity. One standing wave with a node at the atom's location doesn't participate in the atom-cavity coupling and produces a middle cavity peak at $\omega=\omega_{c}$. The other two side peaks are the standard Rabi little peaks, which originate from the interaction of other standing waves in the cavity, which have anti-node at the atomic location. Waveguide affects the Rabi splitting ($\omega_{+}-\omega_{-}$) and modifies it to the value $2\sqrt{2g^{2}-\Gamma^{2}}$. Since we are working in the parameter regime $2g^{2}>\Gamma$ therefore one can take $\omega_{+}-\omega_{-}\approx 2\sqrt{2}g$ as we also confirmed from the Fig.~\ref{Fig2}(c).

\hspace{-3.5mm}$\bullet$ {\bf Weak coupling regime with large backscattering}: In this case, we now introduce a backscattering in the weak coupling regime of cQED (see Fig.~\ref{Fig2}(d)). Since all parameters have some non-zero value here (and for the following two cases), unlike the previously discussed three cases, the transmission and reflection don't take a simpler form and are not reported here. In the context of a weak coupling regime, typically cooperativity $\mathcal{C}:=g^2/(2\kappa\gamma)$ is used as a parameter to quantify the degree of strongness or weakness of a cQED platform \cite{kroeze2023high}. We select $\mathcal{C}=0.125$ for this plot and vary the backscattering rate to $\eta=2\Gamma$ (large backscattering). We find that as $\eta$ is increased, the reflection probability at the resonance begins to show a slight increase (reaching $~10\%$) for the largest $\eta$ value of $2\Gamma$. On the other hand, with the same value of $\eta$, the transmission probability manifests a tiny asymmetric splitting around $\Delta=0$ point. 

\hspace{-3.5mm}$\bullet$ {\bf Strong coupling regime with large backscattering}: This case is shown in Fig.~\ref{Fig2}(e). When cavity backscattering is switched on in the strong coupling regime with $\mathcal{C}=50$, we note that both intra-cavity modes mix up, and the three peak profiles observed in the last case don't repeat themselves here. On the contrary, the transmission spectrum shows two asymmetric dips in the present case. We find that at the largest backscattering value of $\eta=2\Gamma$, these asymmetric peaks turn out to be separated by modified Rabi splitting $2\sqrt{2g^{2}-\Gamma^{2}}$ which takes a value of $\sim 6.2 \Gamma$. 

\hspace{-3.5mm}$\bullet$ {\bf Strong coupling with non-zero atom-cavity detuning}: So far, in all previous cases, we have selected cases in which atom and cavity frequencies resonated. Finally, in this case, we go beyond this condition. In Fig.~\ref{Fig2}(f), we plot the transmission and reflection probabilities while varying atom-cavity detuning $\Delta_{ac}=\omega_{c}-\omega_{eg}\equiv\delta$ between $0$ to $5\Gamma$. We select a strong coupling regime similar to the last case with $\mathcal{C}=50$ for the coupling considerations. The cavity backscattering has been fixed to $2\Gamma$. We find that for the large detuning values (say when $\delta=4\Gamma$) since we are working in a regime where $\omega_{c}-\omega_{eg}>>\Gamma$, the atom is weakly coupled with the cavity-waveguide subsystem. Consequently, an atomic transmission dip appears at $\Delta\sim-5\Gamma$. While at $\omega_{c}$, we observe a ``W-like" pattern centered at ~$\Delta\sim0.25$ due to small backscattering between the cavity modes introduced by the atom-cavity coupling.

%%===================================================%%
%%  Sec.IV: Multiple atom-cavity systems             %%
%%===================================================%%
\section{\label{sec:IV} Periodic Jaynes-Cummings Arrays}

%%%%%%%%%%%%%%%%%%%%%%%%%%%%%%%%%%%%%%%%%%%%%%%%%%%%%%%%%%%%%%%
\subsection{\label{sec:IVA} Transfer matrix approach}
Next, we include multiple atom-cavity systems periodically arranged and coupled through the waveguide continua modes (full Jaynes-Cummings array as shown in Fig.~1). We are interested in calculating the transport properties of a single photon as it propagates from the entire chain of atom-cavity systems. To this end, we divide the whole JC array into $N$ blocks such that each block consists of a single atom-cavity system and a fiber of length $L_{i}$ between any $i+1$ and $i$th atom-cavity system. The transport of photon from the first block can be conveniently described in terms of an atom-cavity transfer matrix ${\bf T}_{1}^{(AC)}$ and a waveguide transfer matrix ${\bf T}^{(F)}_{1}$. For systems with two inputs and outputs, the generic form of atom-cavity transfer matrix under time-reversal symmetry condition can be worked out to be \cite{haus1984waves,fan2002sharp,shen2005coherent}:
\begin{equation}
{\bf T}_{1}^{\rm (AC)}=
\begin{pmatrix}
    1/t^{\ast}       & -r^{\ast}/t^{\ast}  \\
    -r/t      & 1/t  \\
\end{pmatrix},
\end{equation}
while the single photon after propagating through a fiber length $L_{i}$ accumulates a phase $\varphi$ such that the fiber transfer matrix can be represented as
\begin{equation}
{\bf T}^{\rm (F)}_{1}=\begin{pmatrix}
    e^{i\varphi}      & 0  \\
     0    & e^{-i\varphi}  \\
\end{pmatrix}
\end{equation}
Consequently the transfer matrix of the first block ${\bf T}_{1}^{(B)}$ turns out to be:
\begin{equation}
\label{eq:T1B}
{\bf T}_{1}^{\rm (B)}=
\begin{pmatrix}
    e^{i\varphi}/t^{\ast}       & -r^{\ast}e^{-i\varphi}/t^{\ast} \\
    -re^{i\varphi}/t      & e^{-i\varphi}/t  \\
\end{pmatrix},
\end{equation}
while $\varphi\equiv2\pi L/v_g$ with the separation between two consecutive atom-cavity subsystem $L$ is defined through $L=x_2 - x_1$. Since the output from each block serves as the input to the next block in both left and right directions, hence the transfer matrix of the full array ${\bf T}^{(\rm Tot)}$ for $N$ atom-cavity subsystems can be constructed by cascading all block transfer matrices as
\begin{equation}
{\bf T}^{(\rm Tot)}=\prod^{N}_{i=1}{\bf T}^{\rm (AC)}_{i}{\bf T}^{\rm (F)}_{i}=\left({\bf T}^{\rm (B)}_{1}\right)^{N}
\end{equation}
In the last part of the above equation, for the sake of simplicity, we have assumed all atom-cavity systems to be identical, such that all block matrices take the same form.

%%%%%%%%%%%%%%%%%%%%%%%%%%%%%%%%%%%%%%%%%%%%%%%%%%%%%%%%%%%%%%
{\renewcommand{\arraystretch}{1.2}
\begin{table*}
\begin{tabular}{ |p{6.25cm}||p{1.2cm}|p{1.2cm}|p{1.2cm}|p{1.2cm}|p{1.2cm}|p{1.5cm}| }
 \hline
 \multicolumn{6}{|c|}{Table I: Parameters$^\ast$ and their values used in Fig.~\ref{Fig3} and Fig.~\ref{Fig5}.} \\
 \hline
 Regimes & ~~~g & ~~~$\kappa$ & ~~~$\gamma$ & ~~~$\eta$ &  $\Delta_{ac}=\delta$\\
 \hline
Decoupled atoms with no loss   & ~~~0 & ~~~0 & ~~~0 & ~~~1 & ~~~0\\
Decoupled atoms with losses   & ~~~0 & ~~~0.5 & ~~~0.5 & ~~~1 & ~~~0\\
Weak coupling with no backscattering   & ~~~0.25 & ~~~0.5 & ~~~0.5 & ~~~0 & ~~~0\\
Weak coupling with large backscattering   & ~~~0.25 & ~~~0.5 & ~~~0.5 & ~~~2 & ~~~0\\
Strong coupling with large backscattering   & ~~~5 & ~~~0.5 & ~~~0.5 & ~~~2 & ~~~0\\
Strong coupling with atom-cavity detuning  & ~~~5 & ~~~0.5 & ~~~0.5 & ~~~2 & ~~~4\\
 \hline
\end{tabular}
\vspace{-2.5mm}
\caption*{$^\ast$Note that all parameters are defined in units of $\Gamma$.}
\end{table*}
}
\subsection{\label{sec:IVB} A 1D chain of few ($N=2,5,10$) atom-cavity subsystems}
In Fig.~3, we have plotted the transmission spectra of a single photon as it interacts with a JC array consisting of a JC dimer ($N=2$), a JC pentamer ($N=5$), and a JC decamer ($N=10$) under a periodic arrangement. For these 2D plots, we have considered fixed parameters mentioned in Table 1 and plotted $T$ as a function of detuning $\Delta$. For the lattice constant $L$, we have selected a value of $\lambda_0/4$ where $\lambda_0$ is the resonant wavelength (wavelength of the photon emitted by the atom) of the problem. We point out that such lattice constants, in similar waveguide coupled setups, do not produce non-Markovian effects \cite{fang2015waveguide, tufarelli2014non} which is consistent with the model studied in this work. 

\begin{figure*}
\centering
  \begin{tabular}{@{}cccc@{}}
    \includegraphics[width=1.75in, height=1in]{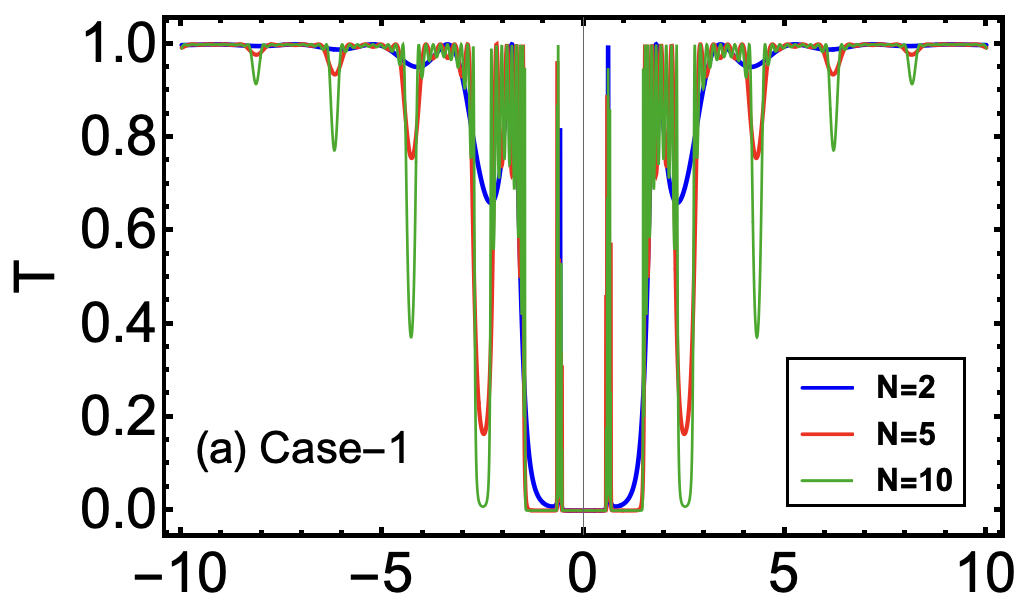} &
    \includegraphics[width=1.55in, height=1in]{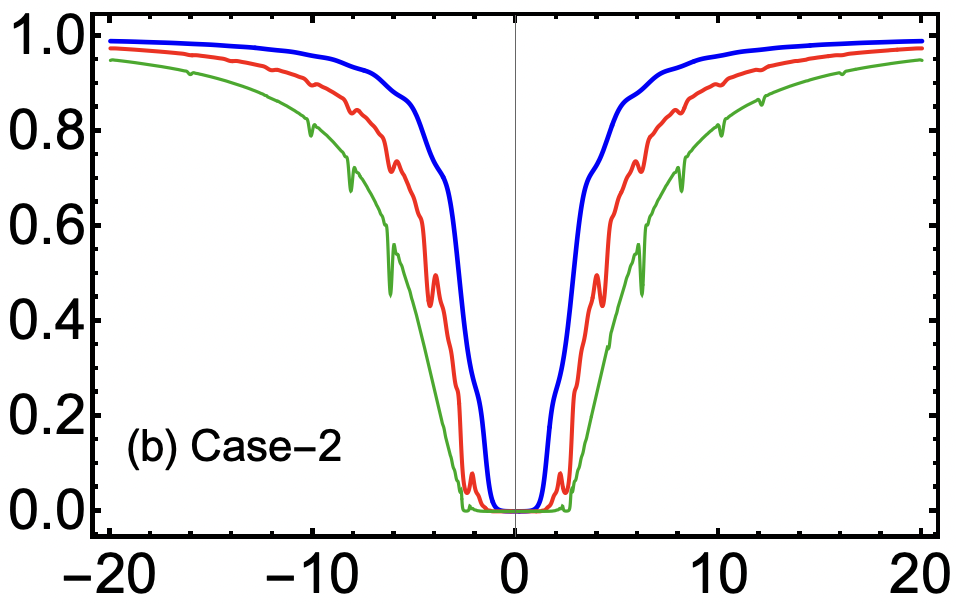} &
    \includegraphics[width=1.55in, height=1in]{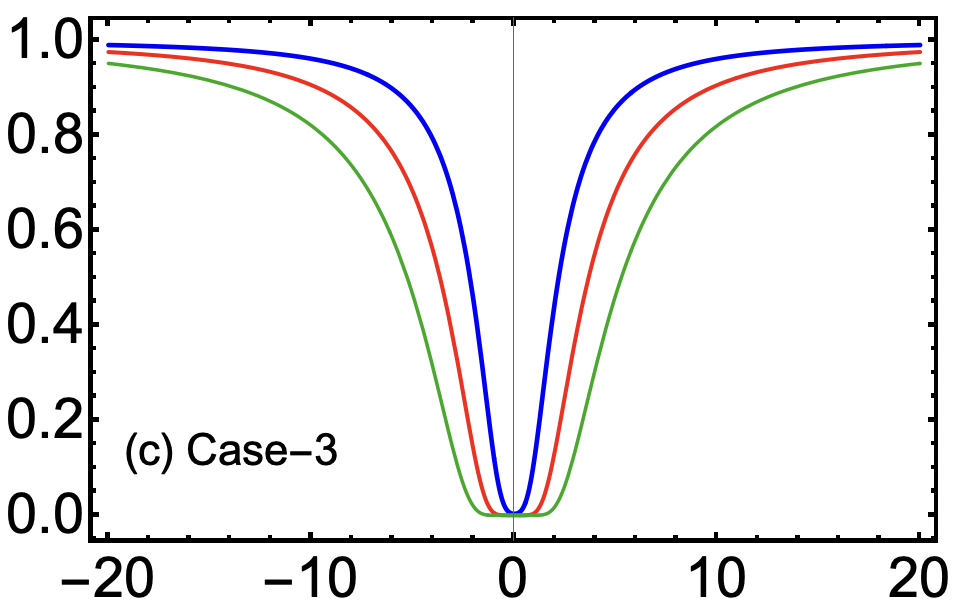} \\
    \includegraphics[width=1.75in, height=1.1in]{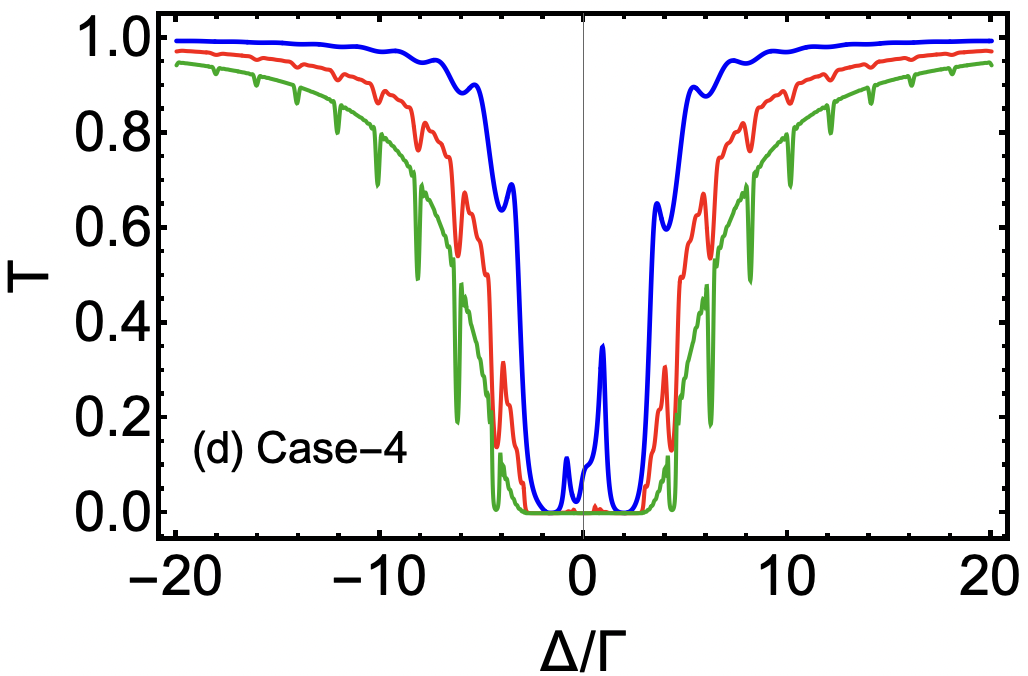} &
    \includegraphics[width=1.55in, height=1.1in]{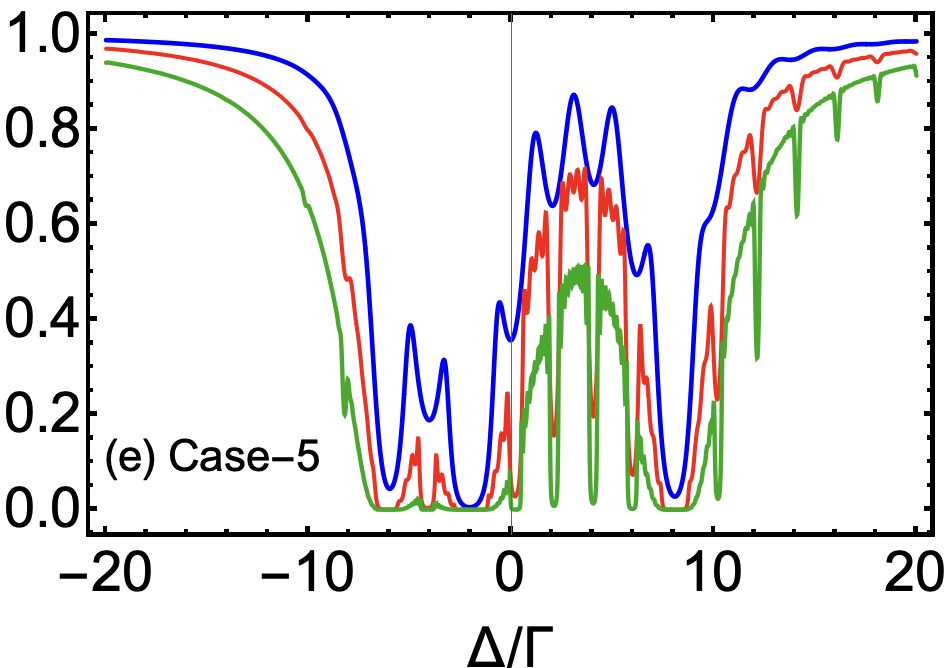} &
    \includegraphics[width=1.55in, height=1.1in]{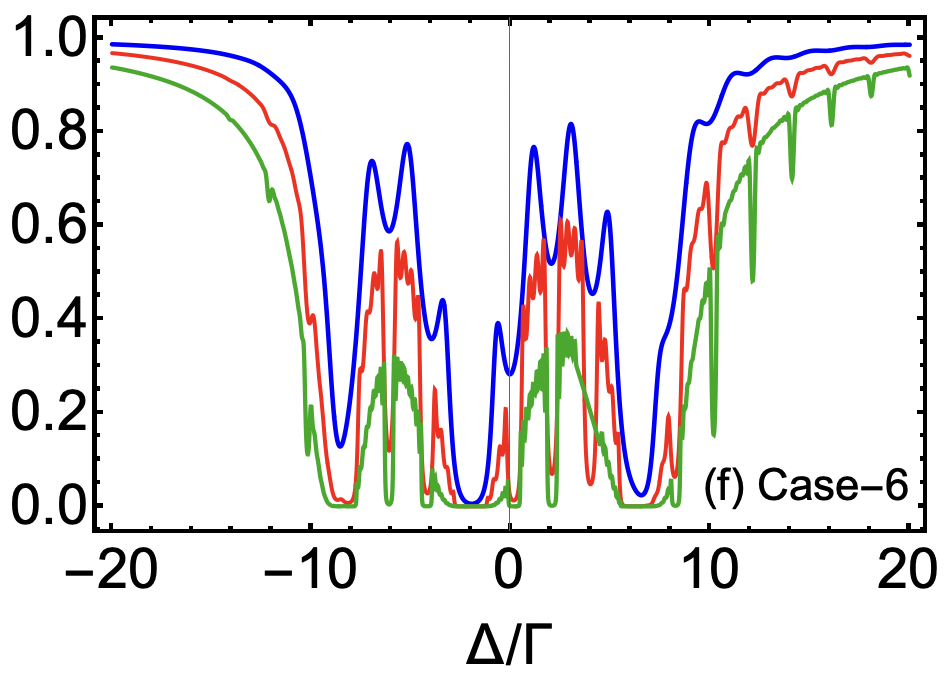} 
  \end{tabular}
  \captionsetup{
  format=plain,
  margin=1em,
 justification=raggedright,
 singlelinecheck=false
}
 \caption{(Color online) Net transmission spectra of a single-photon as it passes through a periodic JC array with two (blue curves), five (red curves), and ten (green curves) atom-cavity subsystem chains. For this plot, we have chosen parameter regimes summarized in Table. 1. For the lattice constant (separation between two consecutive atom-cavity subsystems), we have chosen $L=\lambda_0/4$ where $\lambda_0$ is the resonant wavelength of the problem. Note that the corresponding reflection spectra have not been used to avoid overcrowding of the plots.}\label{Fig3}
\end{figure*}

From the panel of plots in Fig.~\ref{Fig3}, we observe that except for case-3 (weak coupling regime with no backscattering), generally speaking, a band structure begins to emerge as we increase the number of atom-cavity systems in the array. While reaching $N=10$, the bands become most visible, allowing additional frequencies at which the photon can be transmitted through the array. The presence of these sidebands is a known phenomenon in other periodic quantum optical systems \cite{liao2016single, lanuza2022multiband, berndsen2023electromagnetically}, which originates from the constructive and destructive interference among different paths single photon can take while propagating through the whole JC chain. This explanation is evident from our numerical results, where we find that the formation of bands is strongest in the cases when the higher reflection of a single photon is allowed (see, for instance, Fig.~\ref{Fig3}(a), Fig.~\ref{Fig3}(e) and Fig.~\ref{Fig3}(f)). On the other hand, in case 3, since the reflection is minimal due to a weak coupling regime and the absence of backscattering, no bands are formed. 

Additionally (see, for example, from Fig.~\ref{Fig3}(a), Fig.~\ref{Fig3}(b), and Fig.~\ref{Fig3}(c)), we notice that as $N$ is increased, the central locations of transmission dips remain unaffected. Still, the shapes of these resonances have changed into a flat rectangular form at the bottom. Similarly, in the last three cases, the location of the main dips (as observed for $N=1$) remains intact as before, with an additive flatness around the dips formed in the single atom-cavity case.

%%%%%%%%%%%%%%%%%%%%%%%%%%%%%%%%%%%%%%%%%%%%%%%%%%%%%%%%%%%%%%
\subsection{\label{sec:IVC} Extension to infinitely many atom-cavity subsystems and Photonic Band Gaps}
We now turn our attention to an extension of a periodic chain of infinitely many atom-cavity subsystems coupled through a single waveguide. The presence of periodic boundary conditions with a lattice constant $L$ allows us to apply the Bloch's theorem \cite{shen2007stopping, mirza2017chirality, berndsen2023electromagnetically} which results in the following single-photon dispersion relation

\begin{figure*}
\centering
    \includegraphics[width=1.5in, height=2.5in]{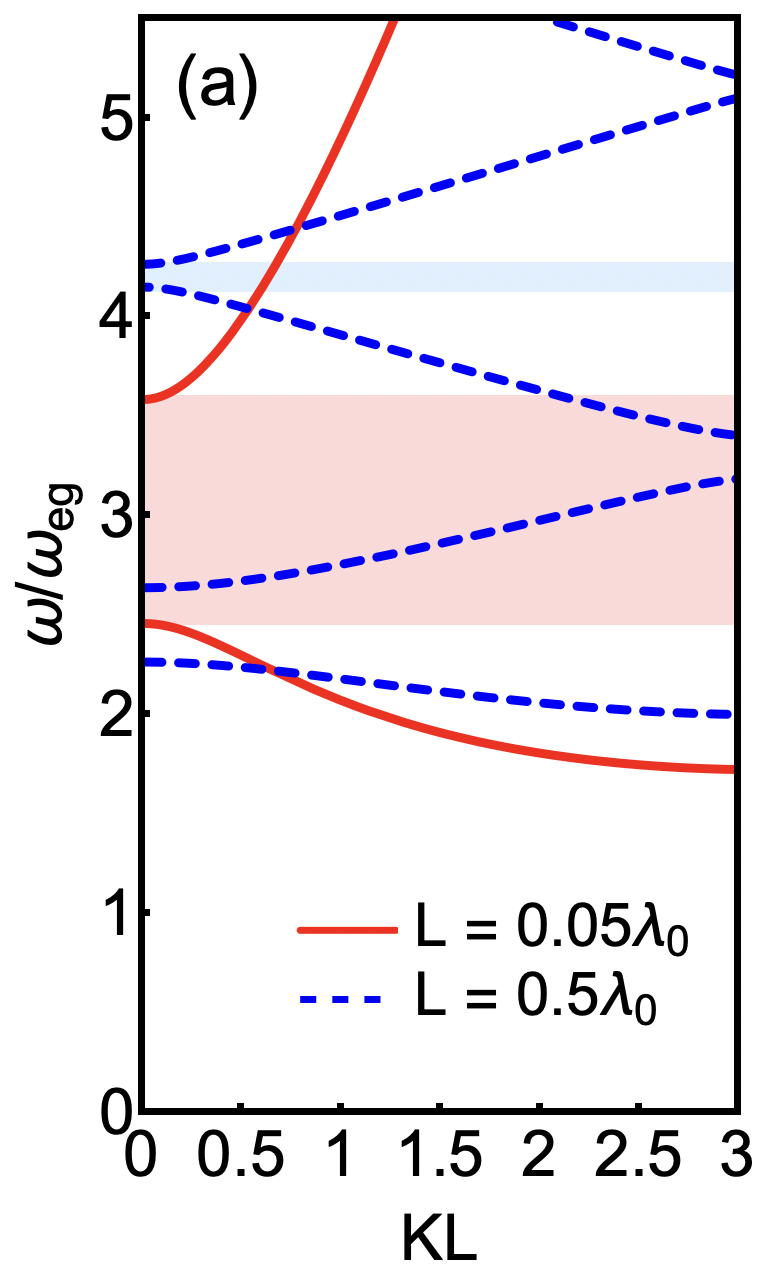} 
    \includegraphics[width=1.35in, height=2.5in]{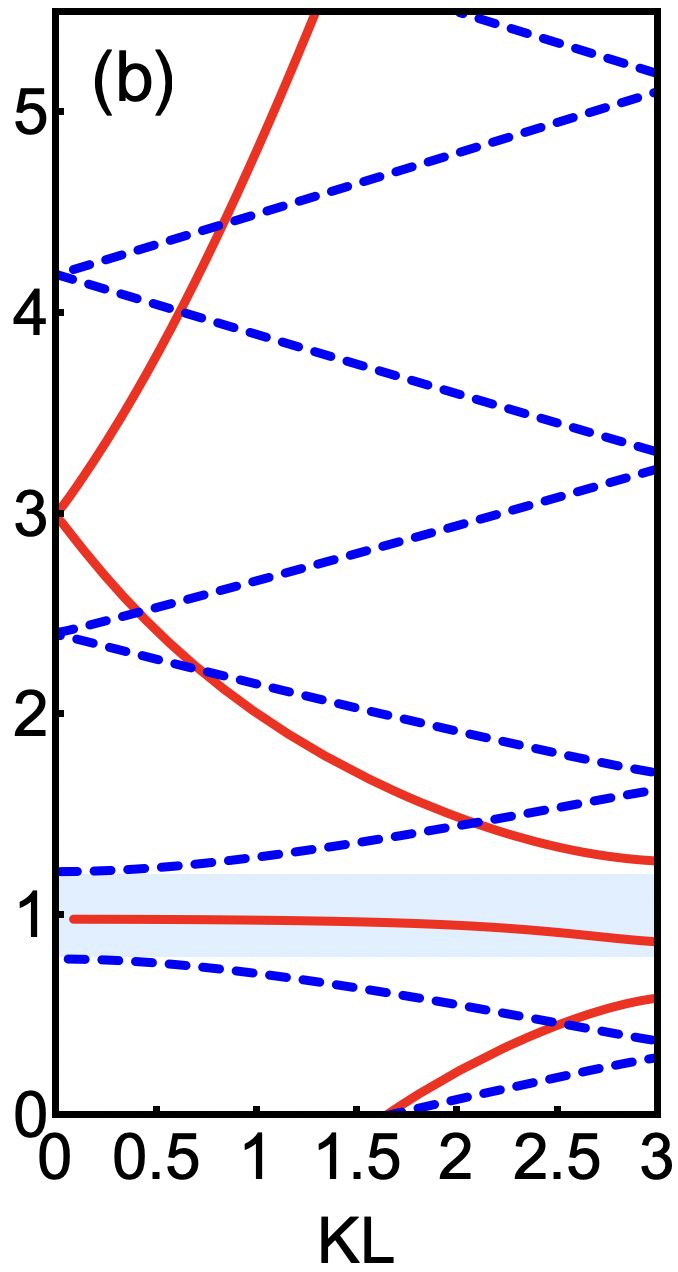} 
    \includegraphics[width=1.35in, height=2.5in]{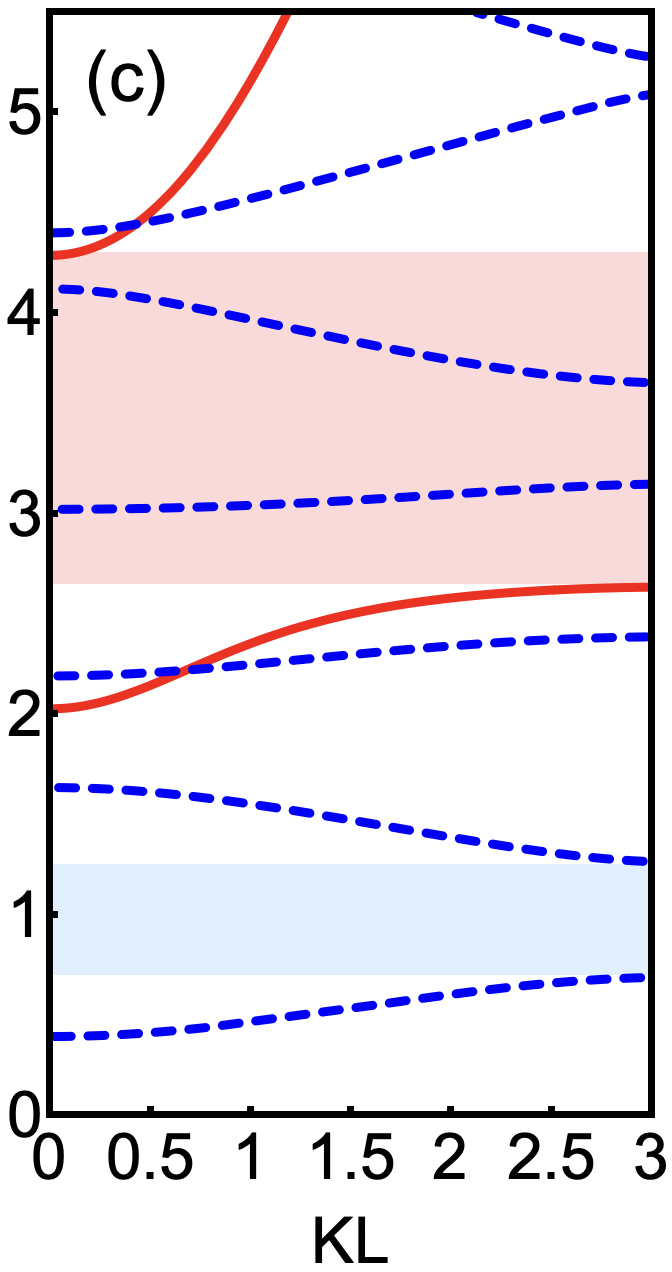} 
    \includegraphics[width=1.35in, height=2.5in]{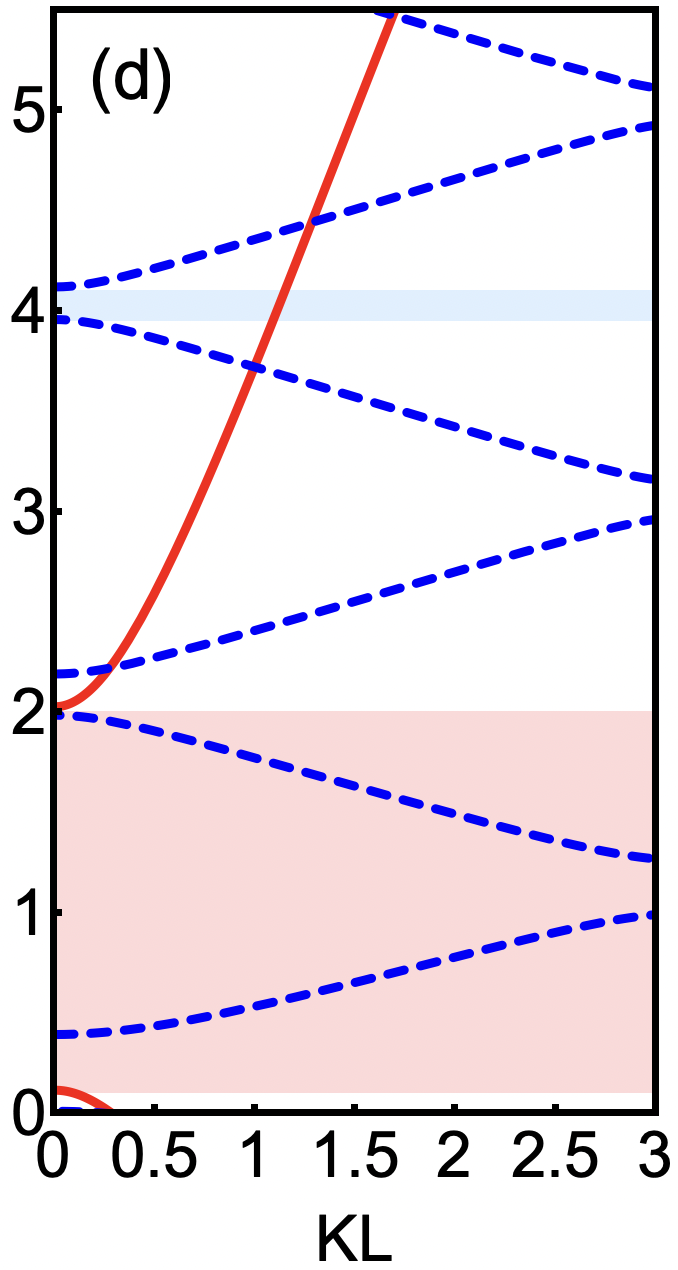} 
    \includegraphics[width=1.35in, height=2.5in]{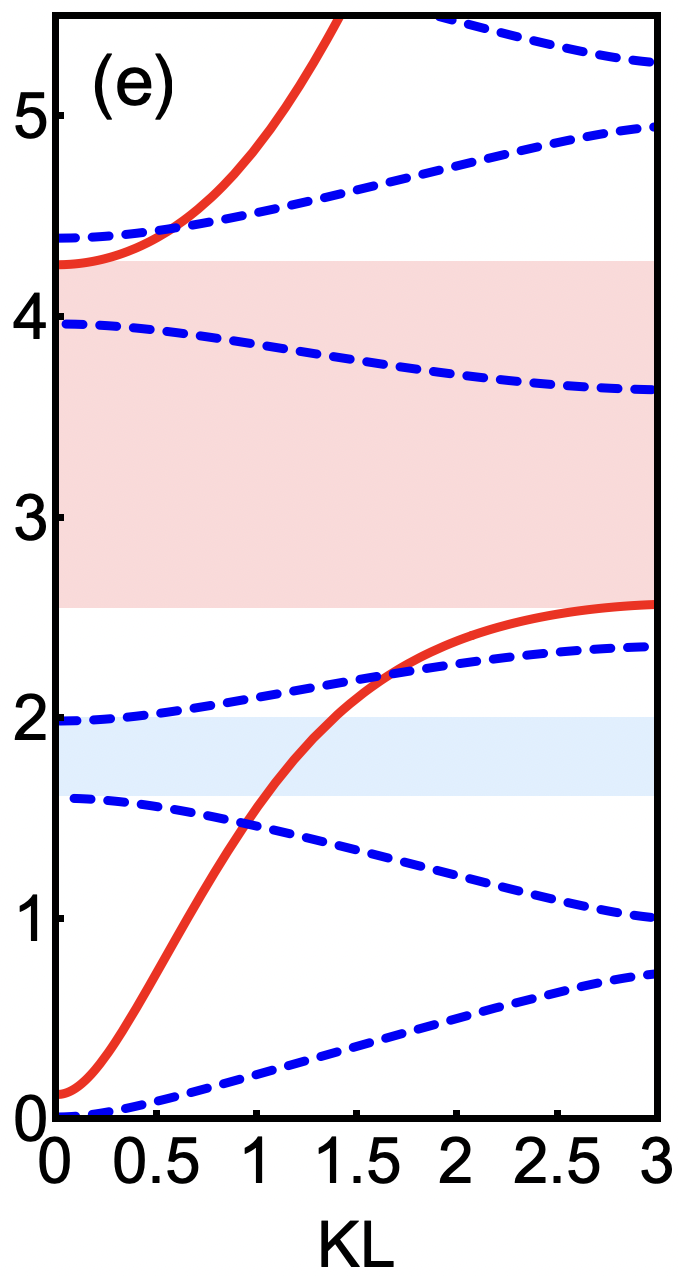} 
  \captionsetup{
  format=plain,
  margin=1em,
 justification=raggedright,
 singlelinecheck=false
}
 \caption{(Color online) Single-photon dispersion curves as it propagates through a infinitely long chain of a JC array. One small (blue dashed curve) and a large (red solid curve) lattice constant has been selected to show the impact of lattice constant on the resulting dispersion properties. In each plot, some of the bands in the small and the large lattice constant case have been colored light red and light blue, respectively, to indicate the presence of gaps. Parameters used in plot: {\bf(a)} $g=0$, $\eta=1$; {\bf(b)} $g=0.25$, $\eta=0$; {\bf(c)} $g=0.25$, $\eta=2$; {\bf(d)} $g=5$, $\eta=2$; and {\bf(e)} $g=5$, $\eta=2$ with $\Delta_{ac}=4$ (note that in previous four cases we have fixed $\Delta_{ac}=0$). In all plots losses have been ignored (i.e. $\kappa=0$, $\gamma=0$) due to validity of application of Bloch's theorem. Furthermore, all parameters are defined in units of $\Gamma$ with the vertical axis label $\omega$ measured in terms of $\omega_{eg}$.}\label{Fig4}
\end{figure*}

\begin{align}
    \cos\left(\mathcal{K}L\right)=\frac{1}{2}\tr\left\lbrace {\bf T}^{(\rm Tot)}\right\rbrace=\mathfrak{Re}\left[\frac{e^{-iqL}}{t}\right],
\end{align}
where $\mathcal{K}$ is the Bloch vector and $q$ is the detuned wavenumber chosen to be defined in terms of atomic transition frequency as $q=(\omega-\omega_{eg})/v_g$. For the present problem the dispersion relation takes the form 
\begin{align}\label{disrel}
   \cos\left(\mathcal{K}L\right)=\frac{\cos(qL)\left(\mathcal{A}\mathcal{C}+\mathcal{B}\mathcal{D}\right)+\sin(qL)\left(\mathcal{A}\mathcal{D}-\mathcal{B}\mathcal{C}\right)}{\mathcal{A}^2+\mathcal{B}^2}. 
\end{align}
Here we have adopted a short notation in which we define
\begin{align*}
    &\mathcal{A}:=-2g^2\left(\Delta+\eta\right)-2\gamma\Delta\kappa -\kappa^2(\delta+\Delta)\\
    &~~~~~~~+(\delta+\Delta)\left\lbrace\Gamma^2+\Delta^2-\eta^2\right\rbrace,\\
    &\mathcal{B}:=-2g^2\kappa+2\kappa\Delta(\delta+\Delta)+\gamma\left\lbrace\Gamma^2+\Delta^2-\eta^2-\kappa^2\right\rbrace,\\
    &\mathcal{C}:=-\Gamma^2(\delta+\Delta)-2\gamma\Delta\kappa-2\Gamma\left\lbrace\gamma\Delta+\kappa(\delta+\Delta)\right\rbrace\\
    &~~~~~~-\kappa^2(\delta+\Delta)+(\Delta+\eta)\left\lbrace-2g^2+(\delta+\Delta)(\Delta-\eta)\right\rbrace,\\
    &\mathcal{D}:=-2g^2(\Gamma+\kappa)+2\Delta(\delta+\Delta)(\Gamma+\kappa)\\
    &~~~~~~-\gamma\left\lbrace-\Delta^2+\eta^2+(\Gamma+\kappa)^2 \right\rbrace,
\end{align*}
note that, similar to Table~1, we have renamed $\Delta_{ac}=\delta$ here to avoid writing subscripts. Equipped with the dispersion relation found in Eq.~(\ref{disrel}), in Fig.~\ref{Fig4}, we plotted the single photon dispersion curves. Since applying Bloch's theorem requires us to consider the no-loss case, these dispersion curves have been plotted for five different cases, as mentioned in the caption of Fig.~\ref{Fig4}. Two different cases of lattice constant (i.e., $L=0.05\lambda_0$ and $L=0.5\lambda_0$) have been considered. Furthermore, in each plot, some band gap regions have been colored (light blue for $L=0.5\lambda_0$ and light red for $L=0.05\lambda_0$) for better visibility.

As some of the critical points, we notice that in each case, whenever a band gap region for the smaller lattice constant was present (as drawn in Fig.~\ref{Fig4}(a), (c), (d), and (e)), it tends to be larger as compared to the large lattice constant band gap. This general feature indicates that smaller lattice constants support enhanced destructive interference, a result which is also known for the waveguide quantum electrodynamics setups (see, for instance, Ref.~\cite{mirza2017chirality, berndsen2023electromagnetically}. 

In this context, the largest band gap for the small lattice constant case is present in Fig.~\ref{Fig4}(d). Interestingly, this plot assumed the highest value of $g=5\Gamma$ (but we cannot call this strong coupling regime of cQED because there are no losses involved to compare $g$ against). Finally, in Fig.~\ref{Fig4}(e), the atom-cavity detuning can be utilized to shift these bands along the vertical scale. Overall, from Fig.~\ref{Fig4}, we conclude that by manipulating parameters $g$, $\eta$, and $\delta$, the regions of forbidden bands for the single photon transport can be engineered.
%%===================================================%%
%%     SectionV: Disordered Case                      %%
%%===================================================%%
\section{\label{sec:V} Disordered Jaynes-Cummings Arrays}
\begin{figure*}[t]
\centering
  \begin{tabular}{@{}cccc@{}}
    \includegraphics[width=2.15in, height=1.43in]{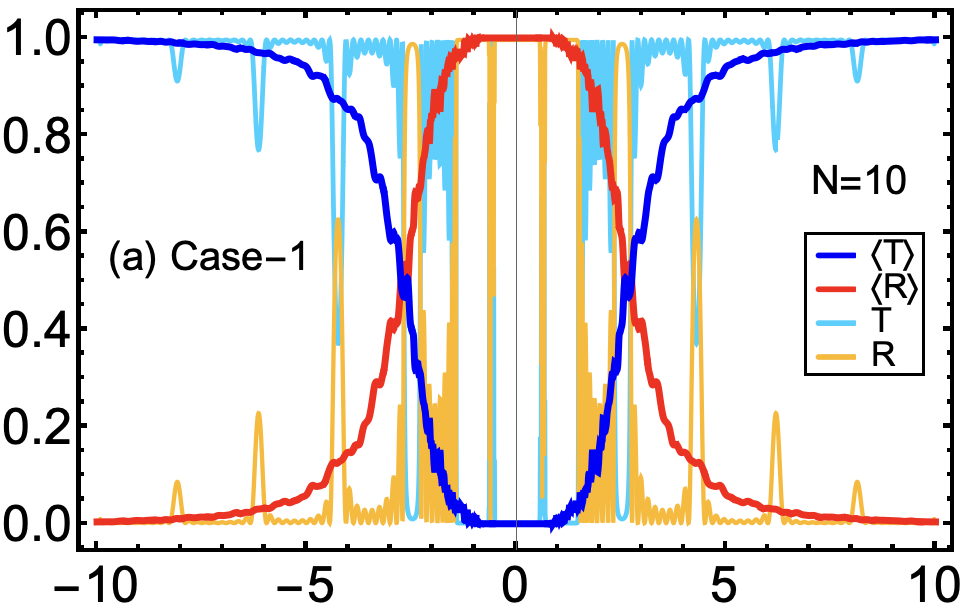} &
    \includegraphics[width=2.15in, height=1.43in]{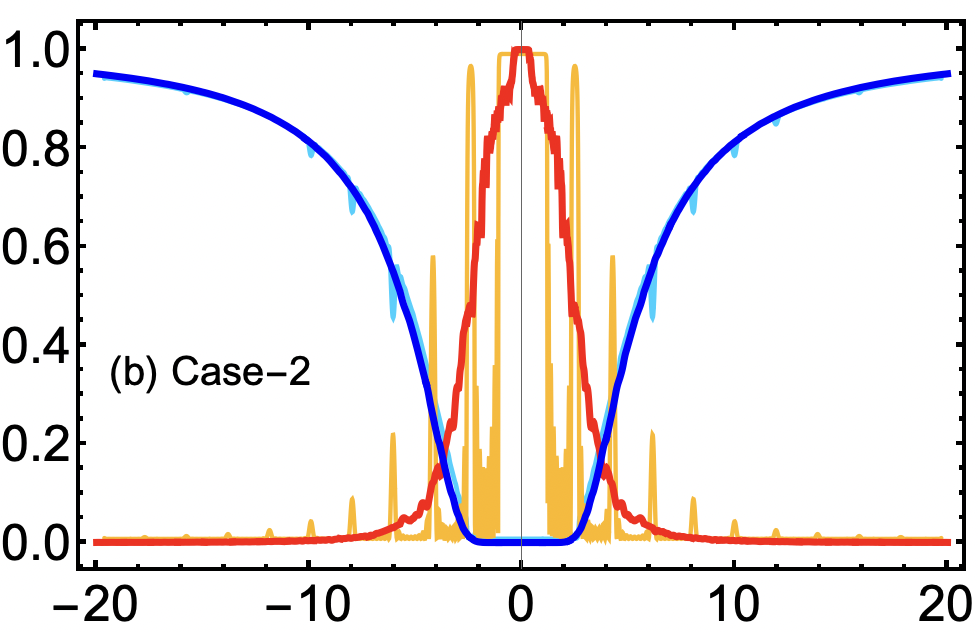} &
    \includegraphics[width=2.15in, height=1.43in]{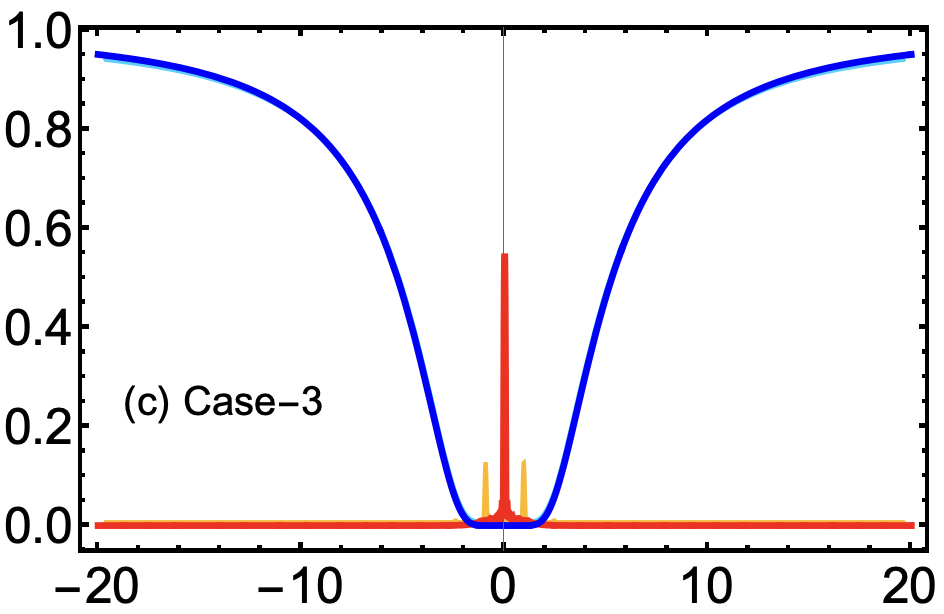} \\
    \includegraphics[width=2.15in, height=1.52in]{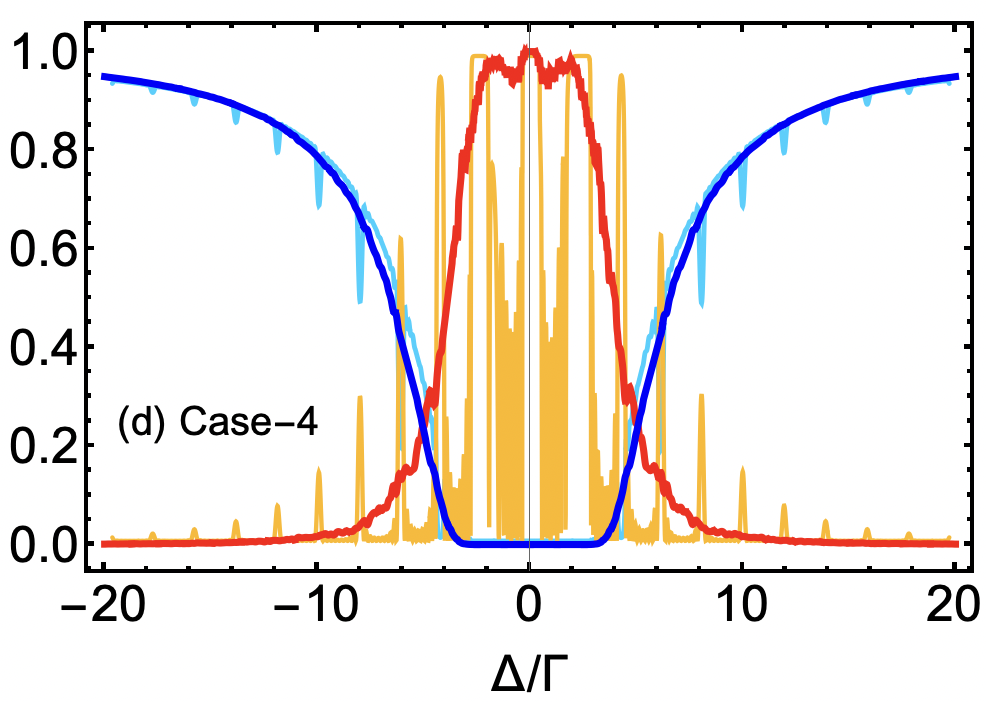} &
    \includegraphics[width=2.15in, height=1.52in]{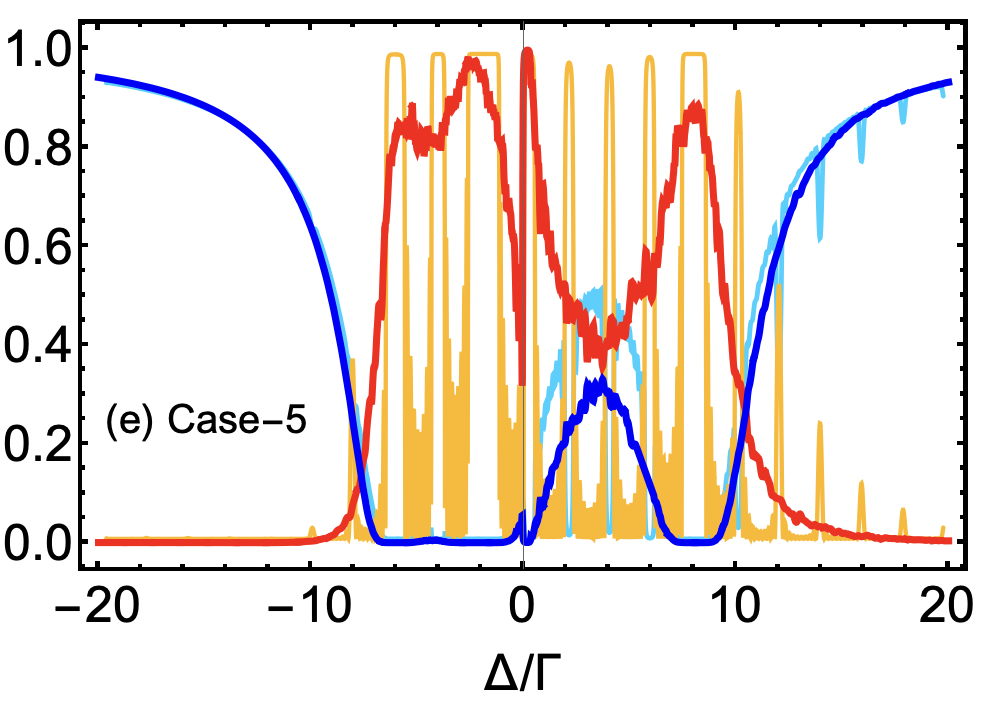} &
    \includegraphics[width=2.15in, height=1.52in]{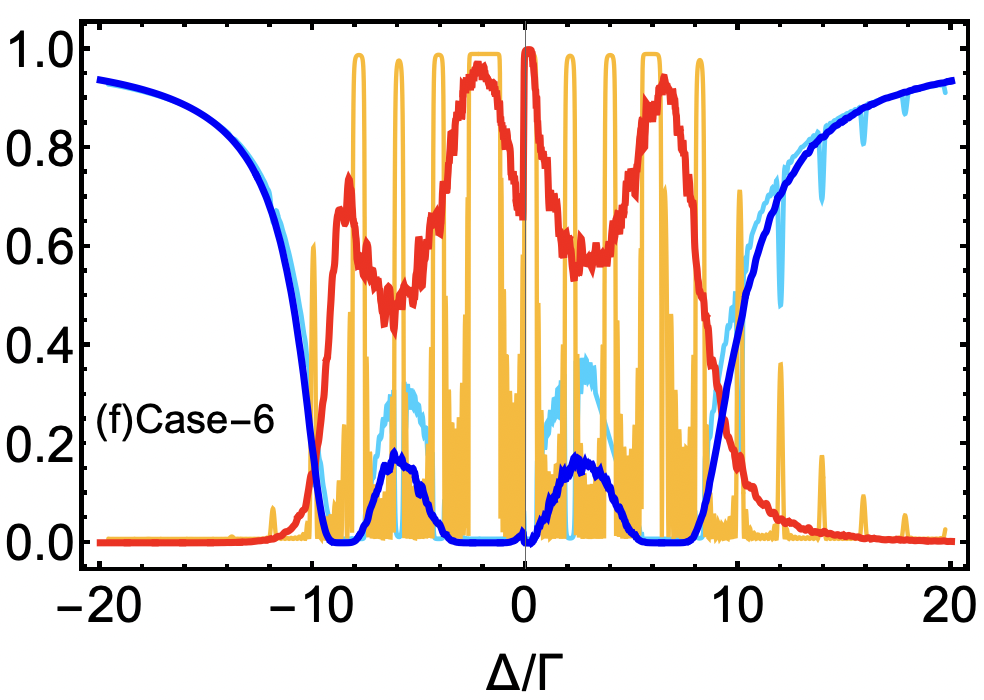} 
  \end{tabular}
  \captionsetup{
  format=plain,
  margin=1em,
 justification=raggedright,
 singlelinecheck=false
}
 \caption{(Color online) Single-photon transport spectra in a waveguide-coupled ten atom-cavity system array. Blue, red, cyan, and orange curves represent the average transmission $\langle T\rangle$, average reflection $\langle R\rangle$, periodic transmission $T$, and periodic reflection $R$ spectra, respectively. The parameters for all six cases have been taken from Table~1. The disordered curves have been averaged over 100 realizations, and the small thickness in $\langle T\rangle$ and $\langle R\rangle$ curves represent the presence of tiny error in the results.}\label{Fig5}
\end{figure*}
Now, we consider the situation of the non-periodic arrangement of atom-cavity subsystems. This section aims to investigate how non-periodicity in the 1D JC lattice affects the overall transport of the single photon. In particular, under what cQED parameter regime can one observe Anderson-like localization \cite{lagendijk2009fifty, segev2013anderson} of the single photons in our problem? By Anderson-like localization, we mean freezing single-photon transport in the JC array due to the destructive interference among the photon's multiple scattering paths. For some recent theoretical and experimental efforts related to photonic Anderson localization, we direct the reader to the references \cite{schwartz2007transport,shapira2005localization,segev2013anderson,skipetrov2014absence,
wiersma1997localization,javadi2014statistical}.

To investigate the localization in a randomly arranged JC array, we have chosen the example of 10 atom-cavity subsystems. The non-periodicity in the arrangement of ring cavities is taken into account through a Gaussian distribution $\mathcal{P}(x)$ of the form
\begin{align}
    \mathcal{P}(x) = \frac{1}{\sqrt{2\pi\sigma^2}}e^{(x-\overline{x})^2/(2\sigma^2)},
\end{align}
where $\overline{x}$ and $\sigma$ represent the distribution's mean and standard deviation, respectively. Like before, the separation between the rings has been measured in units of resonant wavelength $\lambda_0$. The mean location of the ring cavities has been set to their periodic configuration, and disorder of magnitude $\sigma=L/4$ has been chosen. Note that given the geometry of our setup, we have defined $L/4<\sigma\leq L/2$ as the regime of strong disorder while $\sigma\leq L/4$ is the case of weak disorder. Setting the upper bound on $\sigma=L/2$ ensures that the ring cavities don't collide. Furthermore, the radius of each ring, $R$, has been taken to be within the range of $R/\lambda_0 < 0.05$. The final plots are obtained by averaging over one hundred realizations of the same numerical routine where, in each realization, a new set of random arrangements of ring cavities has been utilized due to the aforementioned weak disorder. 

In Fig.~\ref{Fig5}, we present a panel of single-photon transport plots, covering all six cases mentioned in Table~1, but now with the disorder incorporated. Blue and red curves show average transmission $\langle T\rangle$ and reflection $\langle R\rangle$ for the disorder configuration. For comparison, the corresponding transmission (cyan-colored curves) and reflection (orange-colored curves) spectra for the periodic case have also been presented in each plot. First, we notice that due to averaging over 100 trajectories, an error of 10\% is produced, making the disordered curves a bit bolder at certain places (see for instance, Fig.~\ref{Fig5}(e) around $\Delta=5\Gamma$).

Next, as an important result, we find that, besides case-3, in all cases, the disorder tends to smooth out bands formed in the periodic cases. This behavior stems from promoting destructive interferences' among different photon paths due to disorder. Additionally, we find that, around $\Delta=0$ in Fig.~\ref{Fig5}(a)-(d), around $\Delta=-5\Gamma$ in Fig.~\ref{Fig5}(e), and around $\Delta=-1\Gamma$ in Fig.~\ref{Fig5}(f) a range of frequencies become available where $\langle T\rangle\sim 0$. The transmission curve in Fig.~\ref{Fig5}(c) turns out to be immune to a disorder where the periodic and disordered behavior exactly coincide (that is why there are no cyan curves in this plot). However, the reflection curves (red and orange-colored curves) show subtle differences around the $\Delta=0$ region where, in the periodic case, we observe tiny sidebands around the resonant point, which are smooth out in the disorder problem. We explain this behavior by noting that the back reflecting channel has been completely blocked in this case, and the atom is weakly coupled with the cavity. Consequently, the difference between the reflection spectra in periodic and disordered cases is minimal in this scenario. 

\section{\label{sec:VI} Conclusions and Future Directions}
This paper has analyzed the problem of single-photon transport in fiber-coupled multiple atom-cavity systems, as well known as a JC array. Utilizing the real-space quantization approach, we first focused on six important parameter regimes in studying single-photon transport through a single atom-cavity-fiber system. Next, employing the transfer matrix approach, we extended the analysis to the periodic arrangement of ten atom-cavity subsystems. We presented our results in six cases of interest again where we noticed the appearance of a photonic band structure in most cases. Using Bloch's theorem, we investigated the formation of forbidden bands for single photon transport. We concluded that the atom-cavity cooperativity and detuning could be used to engineer the single photon dispersive properties according to the demand. Finally, we introduced non-periodic arrangements through a Gaussian distribution. We pointed out that the non-periodicity in the JC lattice in terms of a weak position disorder smooths out the photonic band structure while (more or less) keeping intact the dominating spectral features observed in the single atom-cavity case.

The rich architecture of the 1D JC array considered in this work allows one to explore new areas of study. For instance, one can investigate the effect of disorder of the ring-resonators more thoroughly by introducing tunneling effects when the resonators are extremely close to each other. In this context, other disorder parameters (for instance, localization length) can be calculated to quantify the amount of single photon localization in longer ($N>10$) JC arrays. This will allow one to draw phase diagrams where single photons fully localized versus fully de-localized states can be tracked. One can also consider disorder in the location of the atoms (say, produced through imperfect atomic trapping techniques or due to finite temperature effects in the environment) while keeping the ring cavities' location fixed in position. We leave these exciting avenues of investigation as possible future directions of this work.

%%===================================================%%
%%          Article Acknowledgements                  %%
%%===================================================%%
\section*{Acknowledgements}
IMM would like to acknowledge financial support from the NSF Grant \# LEAPS-MPS 2212860 and the Miami University College of Arts and Science \& Physics Department start-up funding.

\bibliographystyle{ieeetr}
\bibliography{paper}

\begin{thebibliography}{10}

\bibitem{thompson2013coupling}
J.~D. Thompson, T.~Tiecke, N.~P. de~Leon, J.~Feist, A.~Akimov, M.~Gullans,
  A.~S. Zibrov, V.~Vuleti{\'c}, and M.~D. Lukin, ``Coupling a single trapped
  atom to a nanoscale optical cavity,'' {\em Science}, vol.~340, no.~6137,
  pp.~1202--1205, 2013.

\bibitem{hartmann2008quantum}
M.~J. Hartmann, F.~G. Brandao, and M.~B. Plenio, ``Quantum many-body phenomena
  in coupled cavity arrays,'' {\em Laser \& Photonics Reviews}, vol.~2, no.~6,
  pp.~527--556, 2008.

\bibitem{cataliotti2001josephson}
F.~S. Cataliotti, S.~Burger, C.~Fort, P.~Maddaloni, F.~Minardi, A.~Trombettoni,
  A.~Smerzi, and M.~Inguscio, ``Josephson junction arrays with bose-einstein
  condensates,'' {\em Science}, vol.~293, no.~5531, pp.~843--846, 2001.

\bibitem{jaksch1998cold}
D.~Jaksch, C.~Bruder, J.~I. Cirac, C.~W. Gardiner, and P.~Zoller, ``Cold
  bosonic atoms in optical lattices,'' {\em Physical Review Letters}, vol.~81,
  no.~15, p.~3108, 1998.

\bibitem{lepert2011arrays}
G.~Lepert, M.~Trupke, M.~J. Hartmann, M.~B. Plenio, and E.~Hinds, ``Arrays of
  waveguide-coupled optical cavities that interact strongly with atoms,'' {\em
  New Journal of Physics}, vol.~13, no.~11, p.~113002, 2011.

\bibitem{ruiz2014spontaneous}
J.~Ruiz-Rivas, E.~del Valle, C.~Gies, P.~Gartner, and M.~J. Hartmann,
  ``Spontaneous collective coherence in driven dissipative cavity arrays,''
  {\em Physical Review A}, vol.~90, no.~3, p.~033808, 2014.

\bibitem{saxena2023realizing}
A.~Saxena, A.~Manna, R.~Trivedi, and A.~Majumdar, ``Realizing tight-binding
  hamiltonians using site-controlled coupled cavity arrays,'' {\em Nature
  Communications}, vol.~14, no.~1, p.~5260, 2023.

\bibitem{schetakis2013frozen}
N.~Schetakis, T.~Grujic, S.~Clark, D.~Jaksch, and D.~Angelakis, ``Frozen
  photons in jaynes--cummings arrays,'' {\em Journal of Physics B: Atomic,
  Molecular and Optical Physics}, vol.~46, no.~22, p.~224025, 2013.

\bibitem{heebner2002slow}
J.~E. Heebner and R.~W. Boyd, ``Slow and stopped light'slow'and'fast'light in
  resonator-coupled waveguides,'' {\em Journal of modern optics}, vol.~49,
  no.~14-15, pp.~2629--2636, 2002.

\bibitem{mirza2013single}
I.~M. Mirza, S.~van Enk, and H.~Kimble, ``Single-photon time-dependent spectra
  in coupled cavity arrays,'' {\em JOSA B}, vol.~30, no.~10, pp.~2640--2649,
  2013.

\bibitem{cirac1997quantum}
J.~I. Cirac, P.~Zoller, H.~J. Kimble, and H.~Mabuchi, ``Quantum state transfer
  and entanglement distribution among distant nodes in a quantum network,''
  {\em Physical Review Letters}, vol.~78, no.~16, p.~3221, 1997.

\bibitem{blais2020quantum}
A.~Blais, S.~M. Girvin, and W.~D. Oliver, ``Quantum information processing and
  quantum optics with circuit quantum electrodynamics,'' {\em Nature Physics},
  vol.~16, no.~3, pp.~247--256, 2020.

\bibitem{mirza2015bi}
I.~M. Mirza, ``Bi-and uni-photon entanglement in two-way cascaded fiber-coupled
  atom--cavity systems,'' {\em Physics Letters A}, vol.~379, no.~28-29,
  pp.~1643--1648, 2015.

\bibitem{bostelmann2023multipartite}
M.~Bostelmann, S.~Wilksen, F.~Lohof, and C.~Gies, ``Multipartite-entanglement
  generation in coupled microcavity arrays,'' {\em Physical Review A},
  vol.~107, no.~3, p.~032417, 2023.

\bibitem{mendoncca2020generation}
J.~Mendon{\c{c}}a, F.~de~Moura, M.~Lyra, and G.~Almeida, ``Generation and
  distribution of atomic entanglement in coupled-cavity arrays,'' {\em Physical
  Review A}, vol.~102, no.~6, p.~062416, 2020.

\bibitem{mirza2022dissipative}
I.~M. Mirza and A.~S. Cruz, ``On the dissipative dynamics of entangled states
  in coupled-cavity quantum electrodynamics arrays,'' {\em JOSA B}, vol.~39,
  no.~1, pp.~177--187, 2022.

\bibitem{stannigel2012driven}
K.~Stannigel, P.~Rabl, and P.~Zoller, ``Driven-dissipative preparation of
  entangled states in cascaded quantum-optical networks,'' {\em New Journal of
  Physics}, vol.~14, no.~6, p.~063014, 2012.

\bibitem{meher2022review}
N.~Meher and S.~Sivakumar, ``A review on quantum information processing in
  cavities,'' {\em The European Physical Journal Plus}, vol.~137, no.~8,
  p.~985, 2022.

\bibitem{baum2022effect}
E.~Baum, A.~Broman, T.~Clarke, N.~C. Costa, J.~Mucciaccio, A.~Yue, Y.~Zhang,
  V.~Norman, J.~Patton, M.~Radulaski, {\em et~al.}, ``Effect of emitters on
  quantum state transfer in coupled cavity arrays,'' {\em Physical Review B},
  vol.~105, no.~19, p.~195429, 2022.

\bibitem{qin2016controllable}
W.~Qin and F.~Nori, ``Controllable single-photon transport between remote
  coupled-cavity arrays,'' {\em Physical Review A}, vol.~93, no.~3, p.~032337,
  2016.

\bibitem{liao2010controlling}
J.-Q. Liao, Z.~Gong, L.~Zhou, Y.-x. Liu, C.~Sun, F.~Nori, {\em et~al.},
  ``Controlling the transport of single photons by tuning the frequency of
  either one or two cavities in an array of coupled cavities,'' {\em Physical
  Review A}, vol.~81, no.~4, p.~042304, 2010.

\bibitem{felicetti2014photon}
S.~Felicetti, G.~Romero, D.~Rossini, R.~Fazio, and E.~Solano, ``Photon transfer
  in ultrastrongly coupled three-cavity arrays,'' {\em Physical Review A},
  vol.~89, no.~1, p.~013853, 2014.

\bibitem{ogden2008dynamics}
C.~Ogden, E.~Irish, and M.~Kim, ``Dynamics in a coupled-cavity array,'' {\em
  Physical Review A}, vol.~78, no.~6, p.~063805, 2008.

\bibitem{cheng2022photonic}
D.~Cheng, W.~Wang, C.~Pan, C.~Hou, S.~Chen, D.~Mihalache, and F.~Baronio,
  ``Photonic rogue waves in a strongly dispersive coupled-cavity array
  involving self-attractive kerr nonlinearity,'' {\em Physical Review A},
  vol.~105, no.~1, p.~013717, 2022.

\bibitem{liu2023realization}
Y.~Liu, Z.~Wang, P.~Yang, Q.~Wang, Q.~Fan, S.~Guan, G.~Li, P.~Zhang, and
  T.~Zhang, ``Realization of strong coupling between deterministic single-atom
  arrays and a high-finesse miniature optical cavity,'' {\em Physical Review
  Letters}, vol.~130, no.~17, p.~173601, 2023.

\bibitem{notomi2008large}
M.~Notomi, E.~Kuramochi, and T.~Tanabe, ``Large-scale arrays of ultrahigh-q
  coupled nanocavities,'' {\em Nature photonics}, vol.~2, no.~12, pp.~741--747,
  2008.

\bibitem{tang2022nonreciprocal}
J.-S. Tang, W.~Nie, L.~Tang, M.~Chen, X.~Su, Y.~Lu, F.~Nori, and K.~Xia,
  ``Nonreciprocal single-photon band structure,'' {\em Physical Review
  Letters}, vol.~128, no.~20, p.~203602, 2022.

\bibitem{berndsen2023electromagnetically}
T.~Berndsen and I.~M. Mirza, ``Electromagnetically induced transparency in
  many-emitter waveguide quantum electrodynamics: Linear versus nonlinear
  waveguide dispersions,'' {\em Physical Review A}, vol.~108, no.~6, p.~063702,
  2023.

\bibitem{sheremet2023waveguide}
A.~S. Sheremet, M.~I. Petrov, I.~V. Iorsh, A.~V. Poshakinskiy, and A.~N.
  Poddubny, ``Waveguide quantum electrodynamics: collective radiance and
  photon-photon correlations,'' {\em Reviews of Modern Physics}, vol.~95,
  no.~1, p.~015002, 2023.

\bibitem{anderson1961localized}
P.~W. Anderson, ``Localized magnetic states in metals,'' {\em Physical Review},
  vol.~124, no.~1, p.~41, 1961.

\bibitem{wiegmann1983exact}
P.~Wiegmann and A.~Tsvelick, ``Exact solution of the anderson model: I,'' {\em
  Journal of Physics C: Solid State Physics}, vol.~16, no.~12, p.~2281, 1983.

\bibitem{shen2005coherent}
J.-T. Shen and S.~Fan, ``Coherent single photon transport in a one-dimensional
  waveguide coupled with superconducting quantum bits,'' {\em {Physical
  {R}eview {L}etters}}, vol.~95, no.~21, p.~213001, 2005.

\bibitem{shi2011two}
T.~Shi, S.~Fan, C.~Sun, {\em et~al.}, ``Two-photon transport in a waveguide
  coupled to a cavity in a two-level system,'' {\em Physical Review A},
  vol.~84, no.~6, p.~063803, 2011.

\bibitem{shen2009theoryI}
J.-T. Shen, S.~Fan, {\em et~al.}, ``Theory of single-photon transport in a
  single-mode waveguide. {I}. coupling to a cavity containing a two-level
  atom,'' {\em Physical Review A}, vol.~79, no.~2, p.~023837, 2009.

\bibitem{shen2009theoryII}
J.-T. Shen, S.~Fan, {\em et~al.}, ``Theory of single-photon transport in a
  single-mode waveguide. ii. coupling to a whispering-gallery resonator
  containing a two-level atom,'' {\em Physical Review A}, vol.~79, no.~2,
  p.~023838, 2009.

\bibitem{chen2014scattering}
G.-Y. Chen, M.-H. Liu, and Y.-N. Chen, ``Scattering of microwave photons in
  superconducting transmission-line resonators coupled to charge qubits,'' {\em
  Physical Review A}, vol.~89, no.~5, p.~053802, 2014.

\bibitem{ren2013single}
X.-X. Ren, H.-K. Li, M.-Y. Yan, Y.-C. Liu, Y.-F. Xiao, and Q.~Gong,
  ``Single-photon transport and mechanical noon-state generation in microcavity
  optomechanics,'' {\em Physical Review A}, vol.~87, no.~3, p.~033807, 2013.

\bibitem{jia2013single}
W.~Jia and Z.~Wang, ``Single-photon transport in a one-dimensional waveguide
  coupling to a hybrid atom-optomechanical system,'' {\em Physical Review A},
  vol.~88, no.~6, p.~063821, 2013.

\bibitem{chen2011surface}
G.-Y. Chen, N.~Lambert, C.-H. Chou, Y.-N. Chen, and F.~Nori, ``Surface plasmons
  in a metal nanowire coupled to colloidal quantum dots: Scattering properties
  and quantum entanglement,'' {\em Physical Review B}, vol.~84, no.~4,
  p.~045310, 2011.

\bibitem{spillane2005ultrahigh}
S.~Spillane, T.~Kippenberg, K.~Vahala, K.~Goh, E.~Wilcut, and H.~Kimble,
  ``Ultrahigh-q toroidal microresonators for cavity quantum electrodynamics,''
  {\em Physical Review A}, vol.~71, no.~1, p.~013817, 2005.

\bibitem{armani2003ultra}
D.~Armani, T.~Kippenberg, S.~Spillane, and K.~Vahala, ``Ultra-high-q toroid
  microcavity on a chip,'' {\em Nature}, vol.~421, no.~6926, pp.~925--928,
  2003.

\bibitem{srinivasan2007mode}
K.~Srinivasan and O.~Painter, ``Mode coupling and cavity--quantum-dot
  interactions in a fiber-coupled microdisk cavity,'' {\em Physical Review A},
  vol.~75, no.~2, p.~023814, 2007.

\bibitem{kimble1998strong}
H.~J. Kimble, ``Strong interactions of single atoms and photons in cavity
  qed,'' {\em Physica Scripta}, vol.~1998, no.~T76, p.~127, 1998.

\bibitem{kroeze2023high}
R.~M. Kroeze, B.~P. Marsh, K.-Y. Lin, J.~Keeling, and B.~L. Lev, ``High
  cooperativity using a confocal-cavity--qed microscope,'' {\em PRX Quantum},
  vol.~4, no.~2, p.~020326, 2023.

\bibitem{haus1984waves}
H.~A. Haus, {\em Waves and Fields in Optoelectronics}.
\newblock Prentice Hall, 1984.

\bibitem{fan2002sharp}
S.~Fan, ``Sharp asymmetric line shapes in side-coupled waveguide-cavity
  systems,'' {\em Applied Physics Letters}, vol.~80, no.~6, pp.~908--910, 2002.

\bibitem{fang2015waveguide}
Y.-L.~L. Fang, H.~U. Baranger, {\em et~al.}, ``Waveguide qed: Power spectra and
  correlations of two photons scattered off multiple distant qubits and a
  mirror,'' {\em Physical Review A}, vol.~91, no.~5, p.~053845, 2015.

\bibitem{tufarelli2014non}
T.~Tufarelli, M.~Kim, and F.~Ciccarello, ``Non-markovianity of a quantum
  emitter in front of a mirror,'' {\em Physical Review A}, vol.~90, no.~1,
  p.~012113, 2014.

\bibitem{liao2016single}
Z.~Liao, H.~Nha, and M.~S. Zubairy, ``Single-photon frequency-comb generation
  in a one-dimensional waveguide coupled to two atomic arrays,'' {\em Physical
  Review A}, vol.~93, no.~3, p.~033851, 2016.

\bibitem{lanuza2022multiband}
A.~Lanuza, J.~Kwon, Y.~Kim, and D.~Schneble, ``Multiband and array effects in
  matter-wave-based waveguide qed,'' {\em Physical Review A}, vol.~105, no.~2,
  p.~023703, 2022.

\bibitem{shen2007stopping}
J.-T. Shen, M.~Povinelli, S.~Sandhu, and S.~Fan, ``Stopping single photons in
  one-dimensional circuit quantum electrodynamics systems,'' {\em Physical
  Review B}, vol.~75, no.~3, p.~035320, 2007.

\bibitem{mirza2017chirality}
I.~M. Mirza, J.~G. Hoskins, and J.~C. Schotland, ``Chirality, band structure,
  and localization in waveguide quantum electrodynamics,'' {\em Physical Review
  A}, vol.~96, no.~5, p.~053804, 2017.

\bibitem{lagendijk2009fifty}
A.~Lagendijk, B.~v. Tiggelen, and D.~S. Wiersma, ``Fifty years of anderson
  localization,'' {\em Physics today}, vol.~62, no.~8, pp.~24--29, 2009.

\bibitem{segev2013anderson}
M.~Segev, Y.~Silberberg, and D.~N. Christodoulides, ``Anderson localization of
  light,'' {\em Nature Photonics}, vol.~7, no.~3, pp.~197--204, 2013.

\bibitem{schwartz2007transport}
T.~Schwartz, G.~Bartal, S.~Fishman, and M.~Segev, ``Transport and anderson
  localization in disordered two-dimensional photonic lattices,'' {\em Nature},
  vol.~446, no.~7131, pp.~52--55, 2007.

\bibitem{shapira2005localization}
O.~Shapira and B.~Fischer, ``Localization of light in a random-grating array in
  a single-mode fiber,'' {\em JOSA B}, vol.~22, no.~12, pp.~2542--2552, 2005.

\bibitem{skipetrov2014absence}
S.~E. Skipetrov and I.~M. Sokolov, ``Absence of anderson localization of light
  in a random ensemble of point scatterers,'' {\em Physical review letters},
  vol.~112, no.~2, p.~023905, 2014.

\bibitem{wiersma1997localization}
D.~S. Wiersma, P.~Bartolini, A.~Lagendijk, and R.~Righini, ``Localization of
  light in a disordered medium,'' {\em Nature}, vol.~390, no.~6661,
  pp.~671--673, 1997.

\bibitem{javadi2014statistical}
A.~Javadi, S.~Maibom, L.~Sapienza, H.~Thyrrestrup, P.~D. Garc{\'\i}a, and
  P.~Lodahl, ``Statistical measurements of quantum emitters coupled to
  anderson-localized modes in disordered photonic-crystal waveguides,'' {\em
  Optics express}, vol.~22, no.~25, pp.~30992--31001, 2014.

\end{thebibliography}
\end{document}